\documentclass[3p]{elsarticle}

\usepackage{todonotes}
\usepackage{url}
\usepackage{caption}
\usepackage{subcaption}
\usepackage{booktabs}
\usepackage{graphicx}
\usepackage{multirow}
\usepackage{amsmath}
\usepackage[capitalise, noabbrev]{cleveref}

\usepackage{algpseudocode}
\usepackage{algorithm}
\usepackage{amsfonts}
\usepackage{mathtools}
\usepackage{amsmath}
\usepackage{tabularx}
\usepackage{graphicx}
\usepackage{float}

\usepackage{tikz}
\usetikzlibrary{shapes,arrows}
\usetikzlibrary{fit}
\usepackage{adjustbox}

\algnewcommand\algorithmicinput{\textbf{Local:}}
\algnewcommand\Local{\item[\algorithmicinput]}
\algnewcommand{\IfThenElse}[3]{
  \State \algorithmicif\ #1\ \algorithmicthen\ #2\ \algorithmicelse\ #3}

\usepackage{nomencl}

\usepackage{xcolor}

\usepackage{booktabs}
\usepackage{tabularx}
\usepackage{makecell}
\usepackage[figuresleft]{rotating}

\makeatletter
\newwrite\@nomenclaturefile
\def\makenomenclatureA{%
  \closeout\@nomenclaturefile
  \openout\@nomenclaturefile=PartA\@outputfileextension
  \def\@nomenclature{%
    \@bsphack
    \begingroup
    \@sanitize
    \@ifnextchar[%
    {\@@@nomenclature}{\@@@nomenclature[\nomprefix]}}%
  \typeout{Writing nomenclature file PartA\@outputfileextension}%
  \let\makenomenclature\@empty}
\def\makenomenclatureB{%
  \closeout\@nomenclaturefile
  \openout\@nomenclaturefile=PartB\@outputfileextension
  \def\@nomlature{
    \@bsphack
    \begingroup
    \@sanitize
    \@ifnextchar[%
    {\@@@nomenclature}{\@@@nomenclature[\nomprefix]}}%
  \typeout{Writing nomenclature file PartB\@outputfileextension}%
  \let\makenomenclature\@empty}
\def\printnomenclatureA{%
  \@ifnextchar[%
    {\@printnomenclatureA}{\@printnomenclatureA[\nomlabelwidth]}}
\def\@printnomenclatureA[#1]{%
  \nom@tempdim#1\relax
  \@input@{PartA\@inputfileextension}}
\def\printnomenclatureB{%
  \@ifnextchar[%
    {\@printnomenclatureB}{\@printnomenclatureB[\nomlabelwidth]}}
\def\@printnomenclatureB[#1]{%
  \nom@tempdim#1\relax
  \@input@{PartB\@inputfileextension}}
\makeatother

\usepackage{etoolbox}
\renewcommand\nomgroup[1]{%
  \item[\bfseries
  \ifstrequal{#1}{A}{Subscripts and Sets}{%
  \ifstrequal{#1}{B}{Parameters}{%
  \ifstrequal{#1}{C}{Variables}{%
  \ifstrequal{#1}{D}{Abbreviations}{%
  \ifstrequal{#1}{E}{Appendices}{}}}}}%
]}

\journal{Elsevier Applied Energy}
\begin{document}

\begin{frontmatter}



\title{Efficient Methods for Approximating the Shapley Value for Asset Sharing in Energy Communities}


\affiliation[inst1]{organization={Intelligent and Autonomous Systems Group, Centrum Wiskunde \& Informatica (CWI)},
            city={Amsterdam},
            country={The Netherlands}}

\affiliation[inst2]{organization={Algorithmics Group, Delft University of Technology (TU Delft)},
            city={Delft},
            country={The Netherlands}}

\affiliation[inst3]{organization={Systems Power \& Energy Research Group, University of Glasgow},
            city={Glasgow},
            country={United Kingdom}}
            
\affiliation[inst4]{organization={Smart System Research Group, Heriot-Watt University},
            city={Edinburgh},
            country={United Kingdom}}

\affiliation[inst5]{organization={Department of Electrical and Computer Engineering, Princeton University},
            city={Princeton, NJ},
            country={USA}}

\author[inst1,inst2]{Sho Cremers\corref{cor1}}
\cortext[cor1]{Corresponding author at: Centrum Wiskunde \& Informatica, Science Park 123, 1098 XG Amsterdam, The Netherlands.}
\ead{sho.cremers@cwi.nl}
\author[inst1,inst2,inst5]{Valentin Robu}
\author[inst1,inst2]{Peter Zhang}
\author[inst3]{Merlinda Andoni}
\author[inst3,inst4]{Sonam Norbu}
\author[inst3]{David Flynn}

\begin{abstract}
With the emergence of energy communities, where a number of prosumers invest in shared generation and storage, the issue of fair allocation of benefits is increasingly important. The Shapley value has attracted increasing interest for redistribution in energy settings - however, computing it exactly is intractable beyond a few dozen prosumers. In this paper, we first conduct a systematic review of the literature on the use of Shapley value in energy-related applications, as well as efforts to compute or approximate it. Next, we formalise the main methods for approximating the Shapley value in community energy settings, and propose a new one, which we call the \emph{stratified expected value} approximation. To compare the performance of these methods, we design a novel method for \emph{exact} Shapley value computation, which can be applied to communities of up to several hundred agents by clustering the prosumers into a smaller number of demand profiles. 
We perform a large-scale experimental comparison of the proposed methods, for communities of up to 200 prosumers, using large-scale, publicly available data from two large-scale energy trials in the UK (UKERC Energy Data Centre, 2017, UK Power Networks Innovation, 2021). Our analysis shows that, as the number of agents in the community increases, the relative difference to the exact Shapley value converges to under 1\% for all the approximation methods considered. In particular, for most experimental scenarios, we show that there is no statistical difference between the newly proposed stratified expected value method and the existing state-of-the-art method that uses adaptive sampling (O’Brien et al., 2015), although the cost of computation for large communities is an order of magnitude lower.
\end{abstract}




\end{frontmatter}



\makenomenclatureA

\nomenclature[A,01]{\(i\)}{For agents (households)}
\nomenclature[A,02]{\(k\)}{For classes (clusters)}
\nomenclature[A,03]{\(\mathcal{N}\)}{Set of agents in the community}
\nomenclature[A,04]{\(\mathcal{S}\)}{For subcoalitions formed by agents in community}
\nomenclature[A,05]{}{}

\nomenclature[B,01]{\(SoC^{\text{max}}\)}{Maximum battery $SoC$ [\%]}
\nomenclature[B,02]{\(SoC^{\text{min}}\)}{Minimum battery $SoC$ [\%]}
\nomenclature[B,03]{\(p^{bat,\text{max}}\)}{Maximum (dis)charging power of battery [kW]}
\nomenclature[B,04]{\(\tau^{s}(t)\)}{Export tariff at time $t$ [pence/kWh]}
\nomenclature[B,05]{\(\tau^{b}(t)\)}{Import tariff at time $t$[pence/kWh]}

\nomenclature[C,01]{\(N\)}{Number of agents}
\nomenclature[C,02]{\(T\)}{Number of time steps}
\nomenclature[C,03]{\(t\)}{For time steps}
\nomenclature[C,04]{\(j\)}{For strata}

\nomenclature[C,06]{\(d_{i}(t)\)}{Power demand of agent $i$ at time $t$ [kW]}
\nomenclature[C,07]{\(d_{\mathcal{N}}(t)\)}{Power demand of community $\mathcal{N}$ at time $t$ [kW]}
\nomenclature[C,08]{\(g(t)\)}{Power generated by community RES (wind turbine) at time $t$ [kW]}
\nomenclature[C,09]{\(p^{\text{bat}}(t)\)}{Power of community battery at time $t$ [kW], negative when charging and positive when discharging}
\nomenclature[C,10]{\(p^{\text{grid}}(t)\)}{Power to/from utility grid at time $t$ [kW], negative when selling and positive when buying}
\nomenclature[C,11]{\(SoC(t)\)}{Battery state of charge at time $t$ [\%]}
\nomenclature[C,12]{\(e^{b}(t)\)}{Imported energy at time $t$ [kWh]}
\nomenclature[C,13]{\(e^{s}(t)\)}{Exported energy at time $t$ [kWh]}
\nomenclature[C,14]{\(c^{\text{grid}}_{T}(\mathcal{N})\)}{Annual cost of community $\mathcal{N}$ importing energy from utility grid, with $T=1$ year [\textsterling]}
\nomenclature[C,15]{\(c^{\text{wind}}_{T}(\mathcal{N})\)}{Annual cost of wind turbine of community $\mathcal{N}$, with $T=1$ year [\textsterling]}
\nomenclature[C,16]{\(c^{\text{bat}}_{T}(\mathcal{N})\)}{Annual cost of battery of community $\mathcal{N}$, with $T=1$ year [\textsterling]}
\nomenclature[C,17]{\(\text{DF}\)}{Depreciation factor of battery}
\nomenclature[C,18]{\(c^{}_{T}(\mathcal{N})\)}{Total annual energy cost of community $\mathcal{N}$, with $T=1$ year [\textsterling]}
\nomenclature[C,19]{\(c(\mathcal{N})\)}{Cost function, equivalent to $c_{T}(\mathcal{N})$}
\nomenclature[C,20]{\(\phi_{i}\)}{Annual cost of agent $i$ according to Shapley value [\textsterling]}

\nomenclature[C,21]{\(MC_{i}\)}{Annual cost of agent $i$ according to last marginal contribution (unnormalised) [\textsterling]}
\nomenclature[C,22]{\(\overline{MC}_i\)}{Annual cost of agent $i$ according to normalised last marginal contribution [\textsterling]}
\nomenclature[C,23]{\(d_{p_{-i}}(t)\)}{Average power demand of the community without agent $i$ at time $t$ [kW]}
\nomenclature[C,24]{\(SEV_{i}\)}{Annual cost of agent $i$ according to stratified expected value (unnormalised) [\textsterling]}
\nomenclature[C,25]{\(\overline{SEV}_i\)}{Annual cost of agent $i$ according to normalised stratified expected value [\textsterling]}
\nomenclature[C,26]{\(M\)}{Number of samples of marginal contributions per agent (for adaptive sampling)}
\nomenclature[C,27]{\(RL_{i}\)}{Annual cost of agent $i$ according to adaptive sampling [\textsterling]}
\nomenclature[C,28]{\(K\)}{Number of classes of unique demands in the community}
\nomenclature[C,29]{\(N_k\)}{Number of agents that belongs to class $k$}
\nomenclature[C,30]{\(P()\)}{Multivariate hypergeometric distribution}
\nomenclature[C,31]{\(\hat\phi_k\)}{Cost redistributed to class $k$ by certain redistribution method ($\overline{MC}_k$, $\overline{SEV}_k$, or $RL_{k}$) [\textsterling]}
\nomenclature[C,32]{\(RD_{\phi}(\hat\phi_k)\)}{Relative difference of a redistributed cost to the Shapley value for class $k$ [\%]}
\nomenclature[C,33]{\(RD_{\phi}(\hat\phi)\)}{ Average relative difference of a redistribution method to the Shapley value [\%]}

\nomenclature[D,01]{DF}{Depreciation Factor}
\nomenclature[D,02]{DoD}{Depth of Discharge}
\nomenclature[D,03]{P2P}{Peer-to-Peer}
\nomenclature[D,04]{RES}{Renewable Energy Source}
\nomenclature[D,05]{SoC}{State of Charge}

\printnomenclatureA
\section{Introduction}
\label{sec:introduction}

Recent years have seen a shift towards decentralized energy systems, in which communities of prosumers (consumers with their own local renewable generation capacity and storage) satisfy more of their own energy needs from renewable energy generated from local sources. A number of regions, such as the European Union~\cite{EURLex5217:online} and the United Kingdom~\cite{gov.uk_2021} are providing supportive regulations to encourage communities of consumers to shift away from the dependence on centralized energy generation, and towards more decentralized and local energy generation and storage systems.

\begin{figure}[ht]
\centering
    \begin{subfigure}[c]{0.49\textwidth}
        \includegraphics[width=7cm, clip=false]{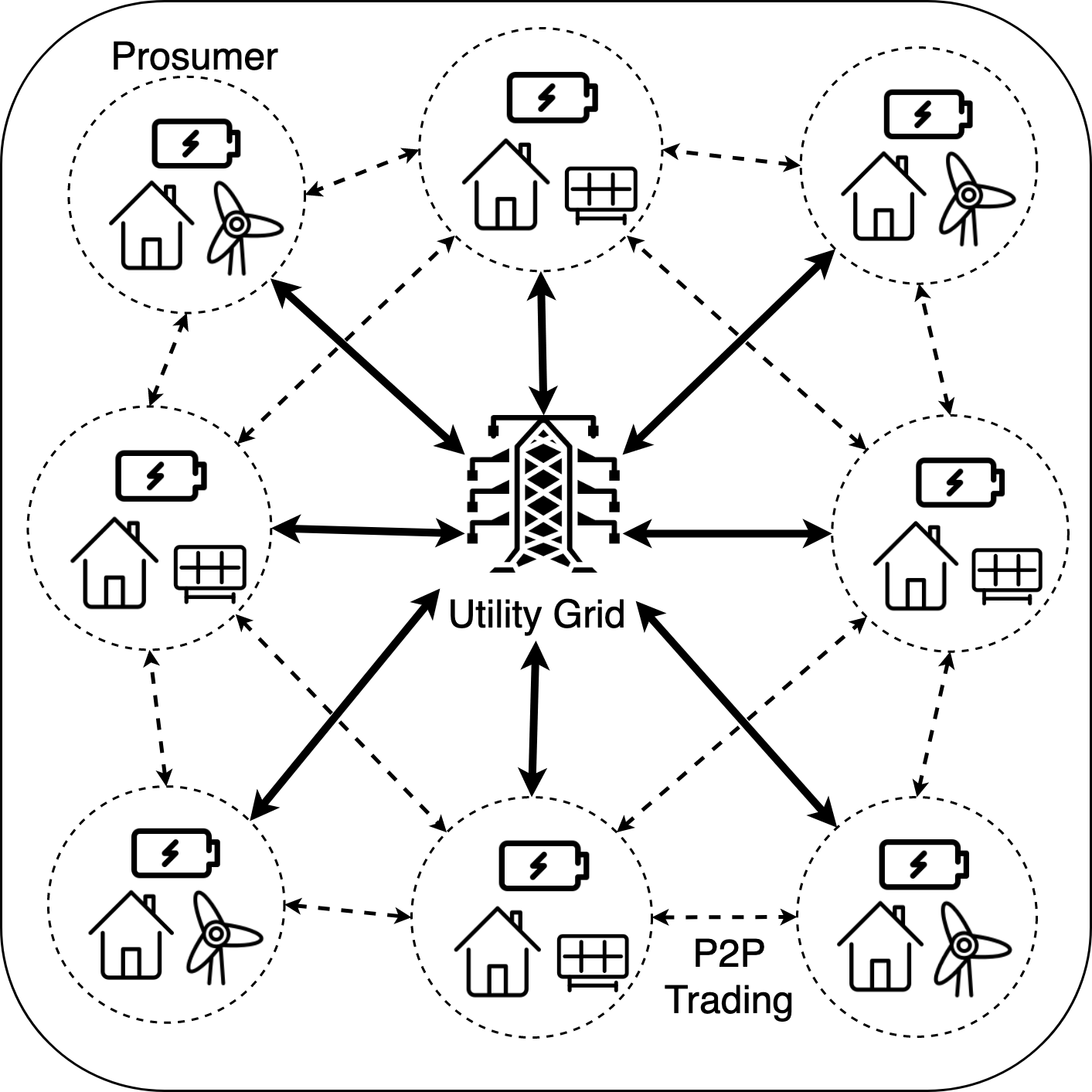}
        \caption{Prosumers with Individually Own Assets}
        \label{fig:individualAssets}
    \end{subfigure}
    ~
    \begin{subfigure}[c]{0.49\textwidth}
        \includegraphics[width=7cm, clip=false]{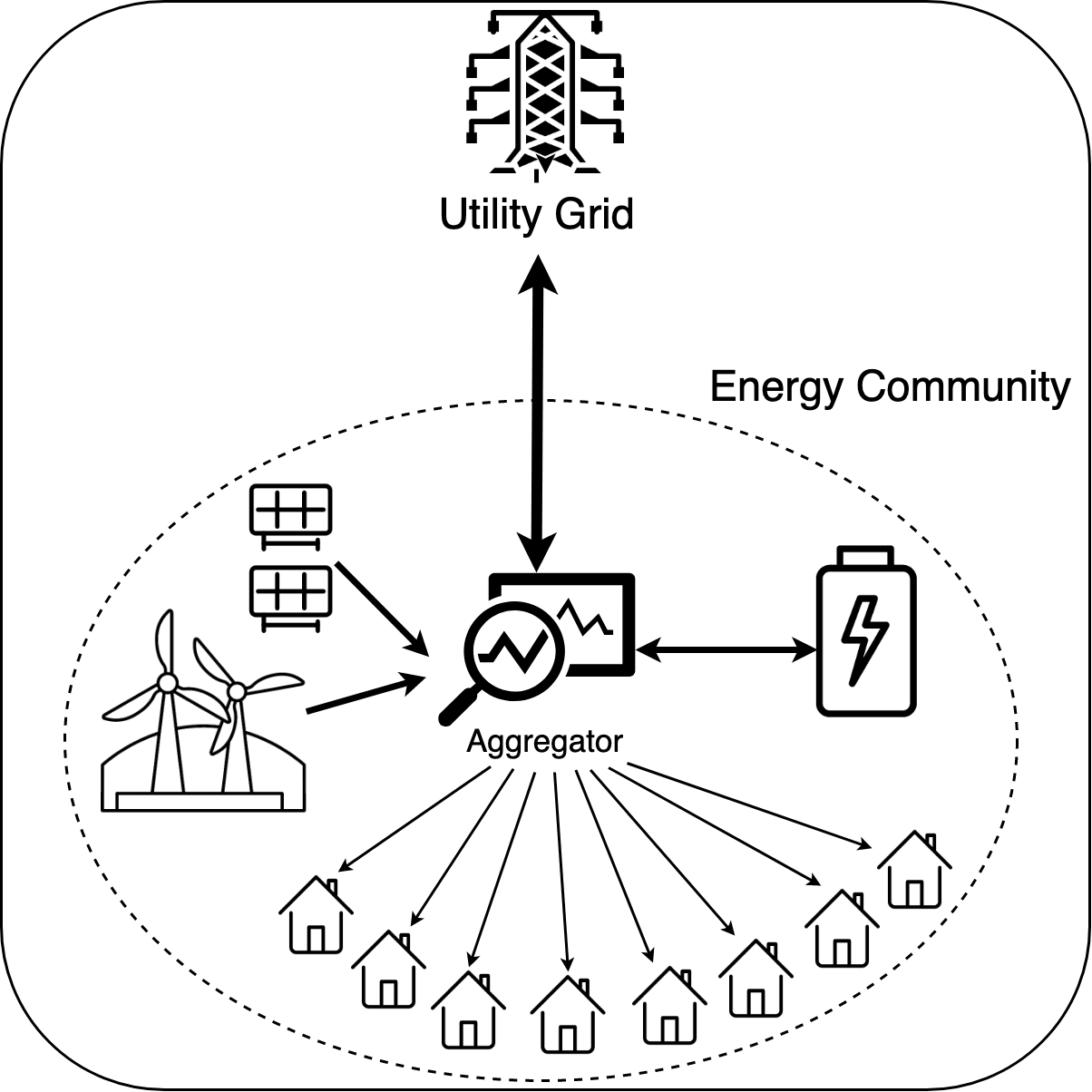} 
        \caption{Energy Community with Shared Assets}
        \label{fig:energyCommunity}
    \end{subfigure}
    \caption{Two different configurations of a community of prosumers, where (a) prosumers with their own energy assets are connected to the central grid individually, and (b) prosumers with jointly-owned assets interact with the grid as a whole through the aggregator.}
    \label{fig:differentCommunities}
\end{figure}

One significant recent trend are transactive energy models that aim to achieve better coordination between production and consumption in local energy systems by use of market-based mechanisms that allow energy exchanges between energy end users and prosumers. In broad lines, there are two main models of organisation for local transactive energy systems~\cite{capper_et_al_2022}. One is peer-to-peer (P2P) energy trading systems, in which prosumers invest in their own energy assets (such as solar PV panels, wind turbine, and or battery storage) and buy and sell energy with their neighbours directly, based on their individually-owned assets~\cite{TUSHAR2021116131,brooklyn_microgrid,ramchurn_AAMAS11,zhang_etal_p2p}. In this scenario, each prosumer is metered separately and pays the value of its \emph{net metered} electricity demand (demand after using its generation and storage capacity). Another is the formation of \emph{energy communities}, where prosumers group together and buy a shared generation resource (such as a community wind turbine) and or a shared community battery. Here, the whole community is \emph{``behind the meter''}, i.e. pays for the net demand of the entire community over the billing period. The differences between the two models are illustrated in \cref{fig:differentCommunities}. Each prosumer in \cref{fig:individualAssets} owns energy sources and a battery, and individually interacts with the central power grid, in which the net demand is counted. It can be seen that the power flow between a prosumer and the utility grid is bidirectional, and any excess generation by the prosumer is sold to the grid. Furthermore, a P2P trading scheme makes buying and selling energy among peers possible, which is represented by dotted arrows in \cref{fig:individualAssets}. On the contrary, the energy community in \cref{fig:energyCommunity} presents a group of consumers sharing energy assets and interacting with the utility grid as a single entity. Net demand is computed for the whole community. These ``behind the meter'' models rely on a \emph{community aggregator}, which controls the energy assets and distributes the generated/discharge power to the households in the community. The aggregator is also in charge of receiving energy from the utility grid whenever there is a deficit and sending back energy to the grid when there is a surplus in the community.

This coalitional model around shared assets is increasingly popular, not just in academic research -- for example, in Scotland, UK, Community Energy  Scotland~\footnote{\url{https://communityenergyscotland.org.uk/}} 
identified 300+ energy communities that formed around a shared energy asset -- typically a wind turbine, but similar examples exist all over the world. Such energy communities can consist of anywhere between several dozens to several hundred houses (e.g. a village or a city neighbourhood), often located on the LV network behind a local transformer. The prosumers share the outputs of jointly-owned energy assets, as well as the energy bill for the aggregate residual demand, i.e. the part of the demand not covered by the local generation and storage assets. Therefore, the community aggregator is not only responsible for the control and distribution of energy in the community but also for allocating any revenues from exporting energy and the bills of the residual demand. Clearly, one of the key challenges in this setting is the redistribution of such costs and benefits to the prosumers in a \emph{fair} way.

Coalitional game theory has long studied such redistribution problems in a wide variety of systems~\cite{Chalkiadakis_al}. A key concept is the Shapley value, first proposed by the Nobel prize-winning economist Lloyd Shapley~\cite{shapley1953value}. The Shapley value has recently begun getting substantial attention in the energy applications -- with a rapid increase in the number of papers using Shapley value in energy systems in recent years (see Section~\ref{sec:relatedWork}). However, a key challenge with the Shapley value is that computing it is exponential in the number of agents
, making exact computation intractable beyond a small number of agents. 

The prior papers dealt with this in several ways. Most consider experimental models with up to a maximum of $\sim$10-20 agents, to keep computations tractable. Another approach is to use some simpler heuristics for cost redistribution (e.g.~\cite{NORBU2021116575}), but it is not clear how close these are to the exact Shapley value. 

Yet another approach is to use sampling. Sampling-based approaches do have merit, and in this paper, we implemented the most advanced sampling-based method we are aware of, that of O'Brien et al.~\cite{Obrien_etal_2015}, which uses reinforcement learning techniques to perform adaptive sampling to calculate the Shapley value. However, they also have disadvantages: for larger settings, a very large number of samples may be needed to get a reasonable approximation of the true Shapley value, which increases the computation cost considerably. Also, in community energy applications, sampling-based methods have the disadvantage that they may not produce a consistent result if they need to be rerun for verification purposes. Community energy schemes rely on the distributed trust of the prosumers in the community, and hence on the ability to sometimes rerun the calculations of the coalition coordinator, if they wish. But, as the calculation at each run depends on which random samples are drawn, results will be slightly different, even on the same data. Hence, there is an important knowledge gap about approximating the Shapley value in larger settings, which our paper aims to address. Specifically, we study both sampling-based and deterministic methods for approximating Shapley, compare their performance w.r.t. the true, exact Shapley value, and derive their computation costs. As one of the contributions of this paper, we introduce a novel redistribution method that approximates the Shapley value well within polynomial time, and compare it to existing methods.

A key open challenge in this space remains determining the ``ground truth'', i.e. computing the exact, true Shapley value to compare other methods to, especially for larger realistically-sized communities (e.g. dozens to several hundred prosumers). Prior approaches, like O'Brien et al.~\cite{Obrien_etal_2015} use a setting of only 20 agents as a ``ground truth'' to compute the exact Shapley, as they naturally find larger settings unfeasible to compute with unique agents. Yet, as we show in our experiments, an approximation method that does poorly for a small number of agents (e.g. 5-20) may actually do well for a realistically sized setting of 100-200 agents. Another important contribution of this paper is that we develop a method to compute the \emph{exact} Shapley value for larger communities, up to 200 agents. Intuitively, the core idea behind the method (see~\cref{sec:kTypesShapley}) is to cluster the agents in a much smaller number of consumption profiles, and use the symmetry of the combinations of agents to greatly reduce the cost of exact Shapley calculation. 

Finally, as part of our contributions, we implemented our method in realistic community case studies, both in terms of demand, generation and battery data used, and in terms of size (up to 200 households), granularity and duration (half-hourly data over a whole year). We used two different datasets, both containing household energy consumption data in the UK and the corresponding wind generation and battery data. One draws data from the Thames Valley vision trial~\cite{TVV} while the other draws data from the Low Carbon London project~\cite{lowCarbonLondon}. This provides a highly realistic case study to provide confidence in the robustness of our experimental comparison results.

The rest of the paper is organised as follows. First, a review of the literature is provided in Section~\ref{sec:relatedWork}. Section~\ref{sec:systemModel} presents the community energy model used. Section~\ref{sec:ShapComputation} introduces the Shapley value and its computation methods. Then, Section~\ref{sec:simulation} presents the experimental comparison across a number of scenarios. Finally, Section~\ref{sec:conclusion} concludes the paper with a discussion.

\section{Literature Study}\label{sec:relatedWork}

This Section presents our systematic study of the literature on Shapley value computation and energy systems.
We note that the distribution of benefits and costs in smart energy systems is a broad one, and Shapley value is just one of the possible solution concepts. It is, however, the most widely used concept and broadly applicable to a variety of settings - with many of the alternatives only applicable in specific settings. Also, Shapley value has a very strong foundation in coalitional game theory, and has had a wide impact in many fields, ever since it was proposed by Nobel-prize winning economist Lloyd Shapley. However, a key problem with applying the Shapley value (especially in the case we study, i.e. energy communities settings with a sizeable number of prosumers) is that it is not computable exactly in settings beyond a few dozen agents, as the computational cost of \emph{exact} Shapley computation is combinatorial. As highlighted by the introduction and our systematic review below, our work helps to close this important knowledge gap by providing and validating computationally-efficient tools to approximate Shapley, with validation in a highly realistic case study of a community energy setting.

The literature study section is divided into two subsections: first, in Section~\ref{sect:lit_Shapley_energy}, we provided a systematic overview of previous works that use Shapley value concept in energy applications, while in Section~\ref{sect:lit_Shapley_aprox} we discuss existing state-of-the-art techniques for Shapley approximation. Our review is enhanced by providing a systematic table that captures and summarises the prior literature related to the application of Shapley values in energy settings, along four key dimensions: the particular sub-area of energy where a referenced paper applied the Shapley value, the computational techniques employed, the type of approach to computing Shapley (whether exact, or approximation based under some assumptions, or both), and finally the number of agents (be it prosumers, households, participants, etc.) that the experimental part of the paper considers. We argue that providing such a table of prior works is important to highlight the current state-of-the-art in the field and make the contribution of this work clearer to the reader.

\subsection{Use of Shapley Values in Energy Applications}
\label{sect:lit_Shapley_energy}

Energy communities are an increasingly important topic of research in energy systems, and a notable number of recent papers consider using the Shapley value as an underlying redistribution method.
Chiş and Koivunen~\cite{Chis_TSG} propose a coalitional cost-game optimisation of a portfolio of energy assets using Shapley value as the underlying redistribution method, modelling a realistic case study of 9 households. Safdarian et al.~\cite{Safdarian_al} use the Shapley value for coalition-based value sharing in energy communities, modelling an energy community in southern Finland with up to 24 apartments.  
Vespermann et al.~\cite{Vesperman_al} study the market design of a local energy community with shared storage and consider a number of solution concepts such as the nucleolus and Shapley values. Their numerical simulations study communities ranging in size from 4 up to 16 prosumers. Robu et al.~\cite{groupbuying_TSG} consider a cooperative coalitional game for energy group buying. While they discuss Shapley value as a solution concept, their focus is on other coalition properties. 

There are also works that study variations of energy communities. \citet{Vinyals_trading} explores a model in which the community consists of prosumers with assets and pure consumers, and the excess energy generated is shared among the community members. Although the work focuses on the energy distribution model that minimises the total cost of the community while meeting regulatory restrictions, it also presents an individually rational cost redistribution scheme. \citet{Long_trading} propose a method for energy trading of excess generation by the prosumers and individual cost calculation based on coalitional game theory and the Shapley value, and tests on a community that consists of 5 prosumers with solar PV generation and energy storage, and 5 consumers with no assets. Similarly, Hupez et al.~\cite{Hupez_2021_EC} compare the use of Nash versus Shapley value concepts in an LV energy community model in which the excess energy of the prosumers is shared among other consumers with a case study of 3 prosumer nodes. 
Singh et al.~\cite{singh_al} present the use of Shapley value for energy trading among microgrids, using a case study of 3 microgrids.
Zhang et al.~\cite{zhang_al} consider the use of Shapley value to divide gains in alliances among retailers in the Chinese energy settlement market, considering alliances up to a size of 9 agents.

\begin{table}[ht!]

\centering
\caption{Summary of published studies on the use of Shapley value in energy applications}
\label{tab:summary}
\resizebox{1.02\linewidth}{!}{
\begin{tabular}{|l|l|l|l|l|}
\toprule
\textbf{Reference} & \textbf{Energy Application Area} & \textbf{Techniques Used} & \parbox[t]{3cm}{\textbf{Type of Shapley\\Computation}} & \textbf{\makecell[tl]{Max No. \\ of Agents}} \\ 
\midrule
Norbu et al.~\cite{NORBU2021116575,norbu_ERCIM,norbu_access} & Energy community& \makecell[tl]{Systematic comparison,\\Data-driven approach} & Marginal contribution & 200 prosumers \\ 
Chiş \& Koivunen~\cite{Chis_TSG} & Energy community & \makecell[tl]{Cost minimisation\\under constraints} & Exact & 9 households \\
Safdarian et al.~\cite{Safdarian_al} & Energy community & K-means clustering & Random sampling & 24 appartments \\ 
\citet{han2019estimation} & Energy community & \makecell[tl]{Sample allocation with\\estimated variance} & \makecell[tl]{Exact\\Stratified sampling} & \makecell[tl]{16 prosumers\\50 prosumers} \\
\makecell[tl]{\citet{kulmala_etal} \&\\\citet{Baranauskas_2022_Shap}} & Energy community & \makecell[tl]{Comparison of cost\\redistribution methods} & Exact \& approximation & \makecell[tl]{6 households} \\
\citet{Long_trading} & Energy community/P2P trading & Fairness analysis & Exact & 10 households \\
\citet{alam_all} & Energy community/P2P trading & \makecell[tl]{Utility maximisation\\under constraints} & \makecell[tl]{Exact\\Random sampling} & \makecell[tl]{10 prosumers\\100 prosumers} \\
\citet{Hupez_2021_EC} & Energy community/P2P trading & \makecell[tl]{Shapley and\\Nash equilibrium analysis}  & Exact & 3 nodes\\ 
Vespermann et al.~\cite{Vesperman_al}& Storage in energy communities & \makecell[tl]{Shapley and\\nucleolus analysis} & Exact & 16 prosumers\\ 
\citet{singh_al} & Trading between microgrids & \makecell[tl]{Decentralised\\coordinated scheduling} & Exact & 3 microgrids\\ 
\citet{Jia_2021_microgrid}  & Trading between microgrids & \makecell[tl]{Cost minimisation\\under constraints} & Exact & 3 microgrids\\
\citet{zhang_al} &  \makecell[tl]{Coordination of electr. retailers} & \makecell[tl]{Transaction cost,\\Resource dependence theory} & Bilateral Shapley value & 9 retailers \\ 
\makecell[tl]{Sharma \& \\Abhyankar~\cite{Sharma_lossAlloc_radial,Sharma_lossAlloc_mesh}} & Loss allocation in distribution &  \makecell[tl]{Exploitation of\\network topology} & \makecell[tl]{Exact\\Sequential Shapley} & \makecell[tl]{24 participants\\68 participants} \\
\citet{Amaris_lossAlloc} & Loss allocation in distribution & \makecell[tl]{Circuit laws,\\Systematic comparison} & Aumann-Shapley value & 35 units \\ 
\makecell[tl]{Pourahmadi \&\\ Dehghanian~\cite{Pourahmadi_lossAlloc}} & Loss allocation in distribution &\makecell[tl]{Benchmarking,\\Systematic comparison} & Exact & 15 units \\ 
\citet{AzadFarsani_lossAlloc} & \makecell[tl]{Allocation of loss reduction \\in distribution} & \makecell[tl]{Point estimation,\\stochastic iterative algorithm} & Exact & 15 units \\ 
\citet{yu2018loss} & \makecell[tl]{Allocation of loss reduction\\in distribution} & \makecell[tl]{Approximation of\\Shapley value and nucleolus} & Aumann-Shapley value & 12 units \\ 
\citet{Vincente-Pastor_2019} & Network coordination & Mechanism design  & Exact & 3 stakeholders\\ 
\citet{Azuatalam_networkCost} & Network cost allocation & SD estimation & Stratified sampling & 25 customers \\ 
O'Brien et al.~\cite{Obrien_etal_2015} & Demand-side response & Reinforcement Learning & Adaptive sampling & 20 participants\\ 
Maleki et. al~\cite{maleki2020shapley} & Coordination of cooling loads & \makecell[tl]{Bounded rational reasoning,\\Dynamic programming} & \makecell[tl]{Bounded rational\\Shapley value} & 15 appartments \\
\citet{Singh_congestion} & Congestion cost allocation & \makecell[tl]{Comparison to Shapley} & Exact & 3 nodes \\ 
\citet{Xiao_congestion} & Congestion cost allocation & Pool-based model & Exact & 6 lines \\ 
\citet{Voswinkel_congestion} & Congestion cost allocation & Constraint optimisation & \makecell[tl]{Exact \& approximation} & 11 congestions \\ 
\citet{cheng_VPP} & Profit distribution in VPP & Coalition and core analysis & Marginal contribution & 3 participants \\ 
\citet{wang_VPP} & Profit distribution in VPP & Real-world feasibility study & Exact & 2 participants \\ 
\makecell[tl]{Dabbagh \& \\Sheikh-El-Eslami~\cite{dabbagh_VPP}} & Profit distribution in VPP & Risk aversion degree & Exact & 6 participants\\ 
\citet{Fang_CHP-VPP} & Profit distribution in CHP-VPP & Particle swarm & Exact (modified) & 4 stakeholders\\ 
\citet{Chattopadhyay_emissionTrading} & \makecell[tl]{Profit distribution\\in emission trading} & Linear programming & Exact & 3 participants\\
\citet{Liao_emissionTrading} & Allocation of emission allowance & Systematic comparison & Exact & 3 power plants\\
\citet{Zhou_2019_carbonObligation} & Carbon obligation allocation & Systematic comparison  & \makecell[tl]{Exact\\Aumann-Shapley value}  & \makecell[tl]{3 DSOs\\20 DSOs}\\ 
\citet{Zhang_emissionAllowance} & Allocation of emission allowance & Entropy, gravity model & Exact & 8 regions\\ 
\citet{Mays_2018_variability}& Net load variability & \makecell[tl]{Consumption behaviour\\profiling}  & Exact & 9 profiles \\
\citet{Zhang_WF_P2G} & \makecell[tl]{Coordinated bidding of wind\\farms and P2G facilities} & \makecell[tl]{Shapley and\\nucleolus analysis}  & Exact & 3 participants\\ 
\citet{Li_NGG-P2G} & \makecell[tl]{Coordinated operation of\\NGG and P2G facilities} & \makecell[tl]{Risk aversion degree, MILP}  & Exact & 2 parties \\
\citet{Churkin_TEP} & Transmission expansion planning & \makecell[tl]{Shapley and\\nucleolus analysis}  & Exact & 5 countries \\
\bottomrule                                                                                  
\end{tabular}}
\end{table}

In addition to the above, applications of Shapley value can be found in many domains within energy systems. The most relevant previous works identified (after a systematic search) on the use of Shapley value in energy application are summarised on \cref{tab:summary}. It reviews 40 selected papers that the authors found to be relevant both to the energy domain and Shapley value computation. It classifies them based on four criteria. The first is regarding the energy application domain in which the Shapley value is applied. The second is techniques used in the work, which could be for computing the Shapley value, but also for solving the underlying problem. The third criterion is how the Shapley value is computed. Most papers compute the exact Shapley value, but there are also many works that make use of approximation methods. Finally, the maximum number of agents used for computing Shapley value in their experimental analyses is given. Some works that apply approximation methods also compute the exact Shapley value as a benchmark. In such cases, the corresponding maximum number of agents for both methods is listed.

From this analysis, we observe that community energy/P2P trading was the most popular application of the Shapley value, but they were also common in other domains, such as the allocation of distribution loss~\cite{Sharma_lossAlloc_radial,Amaris_lossAlloc,Pourahmadi_lossAlloc} and congestion cost~\cite{Singh_congestion,Xiao_congestion,Voswinkel_congestion}, as well as profit distribution in virtual power plants (VPPs)~\cite{cheng_VPP,wang_VPP,dabbagh_VPP}. There were some less obvious applications, namely, cost allocation of net loss variability~\cite{Mays_2018_variability} and coordinated operation of existing facilities and the emerging power-to-gas technology~\cite{Zhang_WF_P2G,Li_NGG-P2G}. 
The range of techniques used by the authors was very broad, ranging from machine learning techniques (e.g., reinforcement learning, K-means clustering), optimisation techniques (e.g., mixed-integer linear programming, particle swarming), to comparison to other allocation methods (e.g., nucleolus, Nash equilibrium).

We found it especially important to provide a classification of Shapley value computation methods and the maximum number of agents considered. Crucially, for exact Shapley computation methods, the number of agents is always kept low to keep the computation tractable, usually to less than ~10 agents. There are few papers that take into account more agents, such as Alam et al.~\cite{alam_all} and previous work by some of the co-authors of this paper (Norbu et al.~\cite{NORBU2021116575,norbu_access}), but these studies do not attempt to compute the Shapley value exactly for a large number of agents and instead use approximation methods like the simple marginal contribution method also used in this study. 
The complexity of computing Shapley often restricts studies to experimental simulations with small numbers of agents - yet, in practice, larger settings appear frequently. 
Realistically sized energy communities have more members, e.g., there are usually 50-200 consumers behind a substation/LV transformer in Europe~\cite{Prettico_all}, or potentially even more sharing an asset such as a large community wind turbine. Hence Shapley approximation methods are needed - yet, the understanding of what is a good approximation for large settings is still lacking. Our work aims to fill this knowledge gap.

\subsection{Approximation Methods for Shapley values}
\label{sect:lit_Shapley_aprox}

Due to the large runtime of the Shapley value computation, it has received strong interest in efficient approximation methods since its introduction. Currently, many approximation methods compute the expected marginal contribution of an agent to the \emph{sampled} coalitions, initially suggested by~\citet{MannShapley1960}. Furthermore, the seminal work of Castro et al.~\cite{castro2009polynomial,castro2017improving} proposes a polynomial calculation method which highlights the concept of \emph{stratified sampling}, which has been refined in other works~\cite{alam_all,han2019estimation}, and is a key concept in the method we develop as well. Many recent works also provide theoretical error bound of sampling-based approximation methods~\cite{Liben-Nowell_FPRAS,maleki2013bounding,aziz2013shapley,BachrachShapleybound,chapman_IJCAI}.

A major obstacle to approximating the Shapley value is that there does not exist a general deterministic approximation method that is a fully polynomial-time approximation scheme (FPTAS), and a fully polynomial-time randomized approximation scheme (FPRAS) is the best one can achieve when approximating the Shapley value~\cite{Liben-Nowell_FPRAS}. Yet, deterministic methods have desirable characteristics for cost redistributions of consumers. Such methods produce the same results after every run given the same inputs, allowing consumers to verify the calculated cost themselves, in contrast to random sampling methods where the redistributed cost can differ depending on the samples drawn. Furthermore, deterministic methods would also guarantee the same cost redistributed to consumers with the exact same demand profile. Such properties can provide consumers with additional trust in the model. \citet{bhagat2014shapley} provides a deterministic Shapley approximation method to their newly proposed budgeted games. Their method is theoretically proven to approximate the Shapley value with a constant additive error by replacing the value function with a relaxed function. However, theoretical analyses of deterministic methods are especially difficult in many real-world energy applications, in which the cost function results from a control procedure over the energy assets over a long time horizon rather than in closed form. Hence, many recent studies in the energy communities have performed empirical analyses to evaluate the performances of the approximation methods (e.g. \cite{Chis_TSG,Safdarian_al,Vesperman_al}).

The publications closest to this work are O'Brien et al.~\cite{Obrien_etal_2015} and Norbu et al.~\cite{NORBU2021116575}. O'Brien et al.~\cite{Obrien_etal_2015} propose an enhancement of the methods first outlined by Castro et al., that uses reinforcement learning to do the stratified sampling in an adaptive way. Their method is one of the methods used as a benchmark in this paper. However, a key limitation of~\cite{Obrien_etal_2015} is that they still use a comparison benchmark of only 20 agents, while we develop a way to compute it exactly for much larger communities. Moreover, we wanted to develop and test some deterministic methods of Shapley value approximation that do not depend on the number of samples and can be reproduced to give the same result. Finally, the work of Norbu et al.~\cite{NORBU2021116575} considers redistribution in realistic community energy settings, starting from marginal value principles, but they do not approximate Shapley value as such. However, with their support, we use the same demand/generation dataset of a community of 200 prosumers in the UK, as it provides a realistic experimental case study to test the methods we develop. Additionally, we also look at a different dataset with a larger number of households to further provide confidence within our methods. This paper is a considerably extended and revised version of preliminary work presented in a poster at the ACM 2022 E-Energy conference~\cite{Cremers_etal_Shapley}.

\section{Community energy model} \label{sec:systemModel}

Consider an energy community $\mathcal{N}$ consisting of a set of $|\mathcal{N}| = N$ prosumers, a shared battery and renewable energy source (RES). In this study, a lithium-ion battery and Enercon E-33 wind turbines~\cite{enercon} with a rated power of 330 kW were considered as the community's energy storage system and RES, respectively. Each prosumer in the community has a half-hourly power demand profile represented as $d_i(t)$ for the power demand of agent $i$ at time step $t$. The final time step of the operation of the system is denoted as $T$. In this study, the data consists of half-hourly demands and generation during a 1~year period, and hence $T = 365 \times 48 = 17520$. 

The demand of the community at time $t$,~$d_{\mathcal{N}}(t)$, is simply the sum of the demands of the agents in the community at $t$, described as the following.
\begin{equation}
d_{\mathcal{N}}(t) = \sum_{i \in \mathcal{N}} d_i(t),\quad \forall t \in \{1, ..., T\}
\end{equation}
Furthermore, a community has a generation profile, $g(t)$, by the jointly owned local renewable energy generation, and the power of the battery,~$p^{\text{bat}}(t)$. The battery is considered charging when $p^{\text{bat}}(t)$ is negative and discharging when $p^{\text{bat}}(t)$ is positive. Finally, a community is required to buy power from the utility grid if the community assets do not provide enough power for the demand. If there is a surplus of power, on the other hand, a community can sell excess power to the grid. The power of the utility grid is denoted as $p^{\text{grid}}(t)$, where the value is positive when power is bought from the grid and negative when power is sold to the grid. 

Given these variables, the following constraint needs to be satisfied at every time step.
\begin{equation}
d_{\mathcal{N}}(t) = p^{\text{grid}}(t) + p^{\text{bat}}(t) + g(t),\quad \forall t \in \{1, ..., T\}
\end{equation}
The constraint assures that the community power demand is met from the power sources. Additionally, when the generation is greater than the demand, all of the energy from the excess power is stored in the battery and or sold to the utility grid.

\subsection{Battery Control Algorithm} \label{subsec:battery_control}
The use of battery was regulated at each time point using the heuristic-based battery control algorithm from~\citet{NORBU2021116575}. The battery keep tracks of its state of charge (SoC), so that the battery capacity is not exceeded. The algorithm first looks at whether the community's demand,~$d_{\mathcal{N}}(t)$, is smaller than the generation of the local RES,~$g(t)$. If the generation is greater than the demand, the battery is charged as long as it has not reached the maximum battery capacity,~$SoC^{\text{max}}$. If the battery has reached the maximum capacity or the surplus power is larger than the maximum (dis)charging power of the battery,~$p^{\text{bat, max}}$, then the remaining energy is sold to the utility grid. The energy is sold with the price of $\tau^{s}(t)$ (pence/kWh), also known as the export tariff. 
If the community power demand is greater than generation, the battery is discharged if it has not reached the minimum battery capacity,~$SoC^{\text{min}}$. If the battery has reached the minimum capacity or the power deficit is larger than $p^{\text{bat, max}}$, then energy is bought from the grid to meet demand. The import tariff, or the price of buying energy from the grid at $t$ (in pence/kWh), is denoted as $\tau^{b}(t)$. More details about the battery control algorithm can be found in~\ref{app:batteryControl}. 

Note that, in the heuristic-based battery control algorithm above, we considered \emph{flat} import/export tariffs (in which the price remains the same throughout the time period of the operation), and moreover, importing or exporting energy to the grid is always worse price-wise than consuming/storing it locally, when possible. This is a realistic assumption in the current climate, when import prices are high, and so-called feed-in tariffs (i.e. tariffs paid to very small renewable generators) are being phased out. 
It is possible to have more advanced control heuristics in case of dynamic or time-of-use prices from the grid that include, e.g. a price prediction component. However, the Shapley computation methods proposed in this paper can also be combined with more complex control cases. This is because the methods we develop apply to the overall cost function, working to minimise the times of iterations needed to recompute it - but are independent of how the control is performed.

\subsection{Community Cost Calculation}\label{subsec:communityCost}

A cost function is a key attribute of a coalitional game. Here, energy cost calculation of the community (or any subset of prosumers) is explained. The community energy cost calculation can be seen as the cost function in this study, and it is required for redistribution methods described in~\cref{sec:redistributionMethods}.

The community energy cost is composed of three components. The first is the cost of energy bought from the grid, subtracted by the revenue of energy sold to the grid during the time period. 
The energy bought and sold at each time point, $e^{b}(t)$ and $e^{s}(t)$ respectively, are determined by the battery control algorithm explained in~\Cref{subsec:battery_control}. The cost $c^{\text{grid}}_T(\mathcal{N})$ is computed as the following.
\begin{equation}
    c^{\text{grid}}_T(\mathcal{N})=\sum_{t=1}^{T} e^{b}(t) \tau^{b}(t)-\sum_{t=1}^{T} e^{s}(t) \tau^{s}(t)
\end{equation}

The second component of the cost for installing and operating the wind turbine,~$c^{\text{wind}}_{T}(\mathcal{N})$. The annual cost is calculated as the following. 
\begin{equation}\label{eq:windTurbineCost}
    c^{\text{wind}}_T(\mathcal{N})=\frac{\text{WT generation capacity} * \text{cost per kW}}{\text{ Lifetime (in years) }}
\end{equation}
The wind turbine generation capacity is calculated as the maximum receiving power from the wind turbine in one time step. The receiving power from the wind turbine was chosen to be $0.006 \times N$ times the power output of one wind turbine. The maximum capacity of the wind turbine increases linearly with the number of prosumers in the community, and hence the cost also increases linearly with the size. The cost of the wind turbine was set to 1072 \textsterling(GB pounds)/kW and lifetime to 20 years, which is realistic for current technologies in the UK market~\cite{NORBU2021116575}.

The last component of the cost is the battery. The cost of the battery, $c^{\text{bat}}_{T}(\mathcal{N})$ is computed as the following.
\begin{equation}\label{eq:batteryCost}
    c^{\text{bat}}_{T}(\mathcal{N})=\frac{\text{battery capacity}* \text{cost per kWh}}{\max \left (\text{ Lifetime (in years) }, \frac{1}{\text{DF}}\right )}
\end{equation}
In this study, the community battery capacity is set to be $5 \times N$ kWh. Similarly to the wind turbine, the battery capacity increases linearly with the community size, and therefore the community battery cost also increases linearly with the community size. The cost of battery per kWh was set to 150 \textsterling/kWh and the lifetime of the battery to 20 years. The variable DF is the depreciation factor of the battery determined by the battery degradation model from~\citet{NORBU2021116575}. Although the battery is given a lifetime, the lifetime can be shortened or additional maintenance costs may be required depending on the number of charge cycles and depth of discharge (DoD). Hence, using a battery degradation model can give a better assessment of the annual battery cost. The details of the battery degradation model are presented in~\ref{app:batteryDegradation}.

The total cost of the community, $c^{}_{T}(\mathcal{N})$ is the sum of the three components, which is the following.
\begin{equation}
\label{eq:CommunityCost}
    c^{}_{T}(\mathcal{N})=c^{\text{grid}}_{T}(\mathcal{N})+ c^{\text{wind}}_{T}(\mathcal{N})+c^{\text{bat}}_{T}(\mathcal{N})
\end{equation}

The community cost can be computed for any subset of agents, and thus the cost contribution of an agent to a group can be determined by comparing the cost of the group with and without the agent. Specifically, every agent in the group contributes equally to the cost of the wind turbine and the battery (from~\cref{eq:windTurbineCost,eq:batteryCost}), but it does not mean the usage of the assets are equal among agents. For example, agents with demand profiles that are well-aligned to the energy generation of the wind turbine will make better use of the community generation assets, resulting in requiring less imported energy from the utility grid to match their demand. On the other hand, agents with demand profiles that are poorly aligned with the generation will put greater pressure on the community battery capacity and equivalently cause more energy to be imported. Therefore, the \emph{marginal value} with which each prosumer causes the total cost to rise is a key factor to consider.

The community energy cost calculation can be seen as a cost function for a set of prosumers with demands. 
The notation of the community cost is simplified to $c(\mathcal{N})$ w.l.o.g., because time horizon $T=1$ year is used to compute costs in the rest of the paper.

\section{Shapley Value Computation}
\label{sec:ShapComputation}
The redistribution of costs or benefits in a game using the Shapley values is considered to be fair in the literature~\cite{shapley1953value,van2002axiomatization}. The cost of prosumer $i$ according to the Shapley value, $\phi_{i}$, is computed as the following.
\begin{equation}
\label{eq:Shapley}
    \phi_{i}=\sum_{\mathcal{S} \subseteq \mathcal{N} \backslash\{i\}} \frac{|\mathcal{S}| !(N-|\mathcal{S}|-1) !}{N !}(c(\mathcal{S} \cup\{i\})-c(\mathcal{S}))
\end{equation}
The marginal contribution of prosumer $i$ to the subcoalition of prosumers $\mathcal{S}$, denoted as $c(\mathcal{S} \cup\{i\})-c(\mathcal{S})$, is how much the prosumer adds to the cost by joining the subcoalition. Then, the Shapley value of agent $i$ can be seen as the mean marginal contribution of $i$ for all possible subcoalitions in the community and all possible permutations of these subcoalitions.

An alternative way to write the Shapley equation that is particularly useful for our approach is through using the concept of \emph{stratum}, given as the following.
\begin{equation}
\label{eq:ShapleyStrata}
     \phi_{i}=\frac{1}{N}\sum_{j=0}^{N-1}\sum_{\substack{\mathcal{S} \subseteq \mathcal{N} \backslash\{i\}, \\ |\mathcal{S}|=j}} \frac{j !(N-1-j) !}{(N-1)!}(c(\mathcal{S} \cup\{i\})-c(\mathcal{S}))
\end{equation}
It can be seen that the marginal contribution of agent $i$ to a subcoalition $\mathcal{S}$ is computed as in~\cref{eq:Shapley}. Then, the marginal contribution is multiplied by the relative frequency of $\mathcal{S}$ in the stratum. A stratum $j$ is a set of all possible subcoalition $\mathcal{S}$ with $|\mathcal{S}| = j$. From this, the expected marginal contribution of agent $i$ to a stratum 0 (empty subcoalition) up to stratum $N-1$ (subcoalition of the whole community except $i$) can be computed. Then, the Shapley value of agent $i$ is equivalent to averaged expected marginal contributions over the strata.

Yet, computing Shapley exactly from these equations has a very large time complexity (exponential to the number of agents in the community, as the marginal contributions to every subcoalition of the community is needed), which makes it intractable very quickly as the community size increases.

\subsection{Methods for Determining Approximate Shapley Values}
\label{sec:redistributionMethods}
In this subsection, we present 3 key methods for computing the Shapley values, starting from the simplest one (last marginal contribution), to increasingly more complex ones such as stratified expected value and adaptive sampling. In Section~\ref{sec:kTypesShapley} we will present a method for exact Shapley computation in the case of a restricted number of types, while in Section~\ref{sec:timeComplexity} we discuss the computational properties of these methods.

\subsubsection{Last Marginal Contribution}
\label{subsec:marginalContribution}

While computing Shapley value directly requires exponential number of steps, it is possible to use the marginal contribution principle to design a much simpler scheme that considers the marginal contribution of each agent w.r.t. the other $N-1$~\cite{kulmala_etal,NORBU2021116575}. Formally, let the cost of an agent $i$ in the community $\mathcal{N}$ be simply the marginal contribution of agent $i$ to the rest of the community, defined as the following.
\begin{equation}
MC_i = c(\mathcal{N}) - c(\mathcal{N}\setminus \{i\})
\label{eq:marginalContribution_new}
\end{equation}

The annual energy cost $MC_i$ of agent $i$ uses the same intuition as~\cref{eq:Shapley} in Shapley value calculation. But, whereas the Shapley value takes the mean marginal contribution of agent $i$ for \emph{every} possible subcoalition in the community, this method computes the cost by only looking at the \emph{last} marginal contribution, making it a much more time-efficient method. However, costs based on the last marginal contributions do not hold the same property of the Shapley values in which the sum of individual cost is equivalent to the total community cost~\cite{shapley1953value}. Hence, the last marginal cost needs to be normalised. The final redistributed cost $\overline{MC}_i$ of agent $i$ according to the normalised last marginal contribution (simply the marginal contribution method from now on) is given as:  
\begin{equation}
\overline{MC}_i = c(\mathcal{N}) \frac{MC_i}{\sum_{q \in \mathcal{N}} MC_q}
\label{eq:marginalContribution_normalized}
\end{equation}

The time complexity of the marginal contribution method is $\mathcal{O}(N)$, so, while simple, it is a very computationally efficient method.

\subsubsection{Stratified Expected Value}
\label{subsec:SEV}

The last marginal contribution method only takes into account the marginal contribution of the last stratum. Starting from this observation, we propose a novel Shapley redistribution scheme that goes a step further and considers the expected marginal contribution for \emph{every stratum}, while still avoiding the huge combinatorial cost of the exact Shapley method. We call this the \emph{stratified expected values} method. 

Formally, for agent $i$, an agent profile $p_{-i}$ that has average energy demands from the rest of the agents in the community is created. The demand of the agent profile $p_{-i}$ at time $t$ is calculated as:
\begin{equation}
    d_{p_{-i}}(t) = \frac{\sum_{q \in \mathcal{N} \setminus \{i\}} d_q(t)}{N-1}, \quad \forall t \in \{1, ..., T\}
\end{equation}

The main idea of the method is that since $p_{-i}$ has the average demand of the rest of the community for every time step, computing the marginal contribution from a set of agents with such a demand profile can approximate the \emph{expected} marginal contribution of that stratum. Since the Shapley value can also be seen as the mean of expected marginal contribution of every stratum, taking the mean of approximated marginal contribution of every stratum should give an average ``in expectation'' value that approximates the Shapley value. Hence, the cost of agent $i$ based on the stratified expected values method $SEV_i$ is calculated as the following.
\begin{equation}\label{eq:stratifiedExpectedValues2}
SEV_i = \frac{1}{N} \sum\limits_{j=0}^{N-1} c(\{1,...,j\} \cup \{i\}) - c(\{1,...,j\}),\quad \text{such that } d_{1} = ... = d_{j} = d_{p_{-i}}
\end{equation}

Similarly to the marginal contribution method, the sum of individual energy costs does not equal the community's total cost since this method uses fictitious agents with demand profiles $d_{p_{-i}}$. Hence, a normalisation step is required, given as follows. 
\begin{equation}
    \overline{SEV}_i = c(\mathcal{N}) \frac{SEV_i}{\sum_{q \in \mathcal{N}}SEV_q}
\label{eq:normStratifiedExpectedValues}
\end{equation}
The time complexity of computing the individual costs with this method is $\mathcal{O}(N^2)$ since for each agent, it requires to calculate the average marginal contribution once for every stratum, ranging from 0 to $N-1$. While this is obviously more than the $\mathcal{O}(N)$ computation of the last marginal value method, it is still much less than the exponential cost of computing the Shapley values, and still very tractable for medium and relatively large community sizes
. The stratified expected value method uses the same intuition as the last marginal contribution method: it considers the last marginal contribution with respect to the expected demand value of the other agents (thus ignoring the combinatorial explosion of computing all orders) - but it does so \emph{for every stratum}, taking an average among them. Thus, it is intuitive to formulate a hypothesis that the stratified expected value method should give a better estimation of the Shapley value than the simpler marginal contribution method, which ignores the strata structure. We explore this hypothesis in \cref{sec:simulation}.

\subsubsection{Adaptive Sampling Shapley Approximation}
\label{subsec:RL}

The previous redistribution methods were deterministic, providing the same numerical results every time the redistributed costs are calculated, given the demands of the agents remain the same. We also compared the performance of a state-of-the-art, random sampling Shapley approximation method. Specifically, we implemented the adaptive sampling method using reinforcement learning introduced by~\citet{Obrien_etal_2015}. For each agent $i$, this method samples a subcoalition randomly from a stratum and computes the marginal contribution of agent $i$ to the subcoalition, repeating this step for $M$ samples predetermined by the user. After every sample, the expected marginal contribution and its estimated standard deviation (SD) of the stratum are updated. The selection of the stratum at the next sample is dependent on the estimated SDs of the strata, where strata with larger spread are more likely chosen. Such a procedure allows to sample more from strata with larger uncertainty, hence sampling more efficiently. Finally, the mean of all expected marginal contributions of the strata is computed as the cost. The details of the algorithm are given in ~\ref{app:RL_approx}.

The time complexity of this redistribution scheme is $\mathcal{O}(N \cdot M)$. Note that, in principle, the number of samples required to approximate the Shapley value well increases faster as the community size increases,  hence $M$ is set to a value that is $M \gg N$. For this study, $M$ was set to 1000 when running this method to assure multiple samples are taken from each stratum.

\subsection{Exact Computation of Shapley Values with K classes} \label{sec:kTypesShapley}

\begin{algorithm*}[!tbp]
  \caption{Create a table containing energy costs of every possible combinations of number of classes}
  \label{alg:CommunityCost}
  \begin{algorithmic}[1]
  \Require{Number of prosumers in the community, $N$. Number of classes, $K$. Number of prosumers in each class, $N_1, N_2, ..., N_K$ with $N_1 + N_2 + ... + N_K = N$ and $N_1 \geq N_2 \geq ... \geq, N_K \geq 1$. Demands of the classes, $d_1, d_2, ..., d_K$, where each $d$ contains half hourly demands during time period of T (1 year).} 
  \Ensure {Table containing costs of all possible subcoalition combinations, $CS$.}
  \Statex
    \Function{CreateTable\_AllSubcoaltionCosts}{$N, N_1, ..., N_K, d_1,...,d_K$}
        \For{all $(n_1, n_2, ..., n_{K}) \in \prod_{k=1}^{K}\{0, 1, ..., N_k\}$} \Comment{Cartesian product}
            \State $\mathcal{S} = \bigcup_{k=1}^{K} \{1,...,n_k\} \;$ with demands $d_k$ \Comment{Union of sets with $1,...,n_k$ having the same demand $d_k$}
            \State {$CS[n_1, n_2, ..., n_{K}] = c(\mathcal{S})$} \Comment{Cost of subcoalition, Eq.~\ref{eq:CommunityCost}}
        \EndFor

        \State \textbf{return} $CS$
    \EndFunction
  \end{algorithmic}
\end{algorithm*}
\begin{algorithm*}[!tbp]
  \caption{Compute Shapley Exact values of $K$ classes}
  \label{alg:shapley}
  \begin{algorithmic}[1]
  \Require{$N$, $N_1, ..., N_K$. Table of energy costs of subcoalitions, $CS$} 
  \Ensure {Array of Shapley values (redistributed cost) for agents in each class, $Shap$}
  \Statex
    \Function{ComputeExactShapley}{$N, N_1, ..., N_K, CS$}
        \For{$k \gets 2$ to $K$}\Comment{Iterate through every class except the first one}
            \State{$Shap_k \gets 0$}
            \For{$j \gets 0$ to $N-1$} \Comment{Iterate through every stratum}
                \For{all $(n_1, n_2, ..., n_{K}) \in \{0, 1, ..., \max(j,N_k - 1)\} \times \prod_{i=1, i\neq k}^{K}\{0, 1, ..., \max(j,N_i)\}$} 
                    \If{$j = \sum_{i=1}^{K}(n_i)$}
                        \State{$rel\_f \gets P(\{n_1,..., n_K\}, \{N_1,...,N_k-1, ..., N_K\}, N, j)$} \Comment{Hypergeometric function}
                        \State{$mc \gets CS[n_1, n_2,...,n_k+1,..., n_K] - CS[n_1, n_2,...,n_k,..., n_K]$}\Comment{Marginal contribution}
                        \State{$Shap_k \gets Shap_k + rel\_f * mc$}
                    \EndIf
                \EndFor
            \EndFor
            \State{$Shap_k \gets \frac{1}{N}Shap_k$}
        \EndFor
        \State{$Shap_1 \gets \frac{1}{N_1}(CS[N_1, N_2, ..., N_K] - \sum_{i=2}^{K} N_i * Shap_i)$}
        \State \textbf{return} $Shap$
    \EndFunction
  \end{algorithmic}
\end{algorithm*}

Given the above redistribution methods, what is really needed is the ``ground truth'' consisting of the exact Shapley values, to compare the performance of these approximation methods for a realistic size community (e.g. $N=200$ prosumers behind a transformer). Prior works that do this, like~\citet{Obrien_etal_2015}, reduce the number of prosumers to $N=20$ to compute the exact Shapley, but we argue this method is not really a satisfactory way to proceed. This is because, crucially, the quality of an approximation for a larger community (e.g. $N=50, 100$ or $200$ agents) can be very different than for a very small number of agents, up to 20 (we clearly show this effect in our experiments as well). 

The key intuition is that, while computing the Shapley values of $N$ unique agents requires a time complexity that is exponential to $N$, the computation time can be significantly reduced if the community consists of a limited number of classes of agents, where agents in the same class have the same demand profile. 

Let the new model be defined as the following. A community $\mathcal{N}$ still consists of $N$ agents, with now $K$ classes of demand profiles in the community such that every agent belongs to one class, and all the agents in the same class have equal half-hourly demands. We assume w.l.o.g. that classes are ordered by size, i.e.
\begin{equation}
    N \geq N_{1} \geq ... \geq N_{K} \geq 1
\end{equation}
where $N_{k}$ is the size of the class $k$. Then, the number of all possible energy costs of subcoalition in the community is $(N_1+1) \times ... \times (N_K+1)$, since from each class $k$ you can have 0 to $N_k$ agents being part of the subcoalition. This is important to note because computing the annual costs of subcoalitions (the cost function) is the most computationally expensive part of computing the redistributed costs, since it has to run through one year of half-hourly demands datapoints every time the cost function is called. The energy costs of every subcoalition may be used multiple times to compute the marginal contribution in Shapley calculation, hence storing the values in a table of the dimension $(N_1+1) \times ... \times (N_K+1)$ can be time-saving. \cref{alg:CommunityCost} shows the creation of the table storing the costs of all possible subcoalitions in a community. 

\begin{figure}[t]
    \centering
    \includegraphics[width=0.5\linewidth]{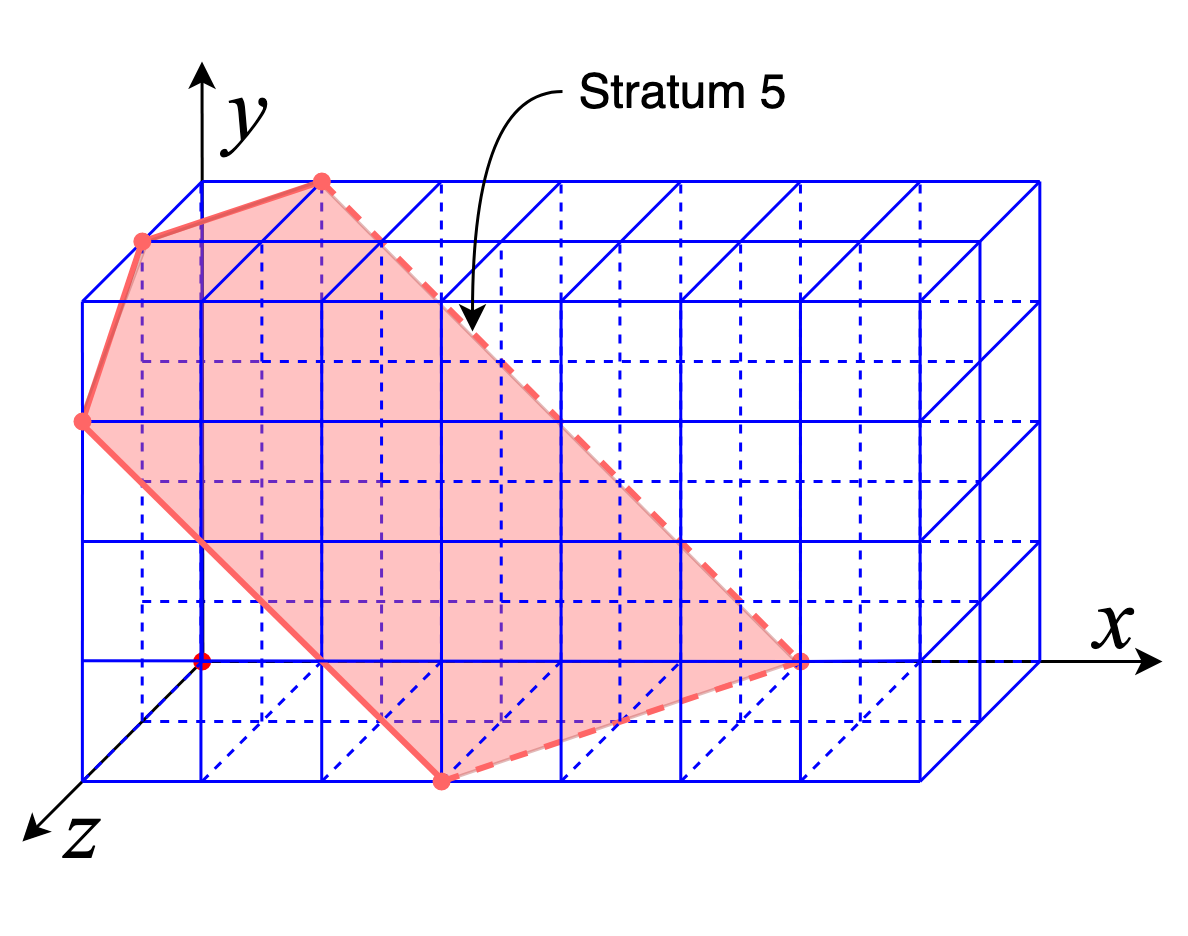}
    \caption{A representation of all possible subcoalitions in a community of $K=3$ classes with $N_1 =7$, $N_2 =4$, and $N_3 =2$. }
    \label{fig:strata}
\end{figure}

The table containing costs of every subcoalition can also be represented as a hyperrectangle of $K$ dimensions and the size $N_1 \times ... \times N_K$. Each axis represents the number of agents in the class. A stratum can be represented in such a hyperrectangle as a hyperplane cutting through in which the sum of axes equals the size of the stratum. Hence, strata correspond to planes parallel to each other. \cref{fig:strata} shows an example case where $K=3$ with $N_1 =7$ (x-axis),  $N_2 =4$ (y-axis), and $N_3 =2$ (z-axis). Stratum 5 is represented by the plane, where $x+y+z=5$. 

Once the energy costs of all possible subcoalitions have been computed, the Shapley values of the agents, which is the energy cost the agent owes to the community, can be found. Because of the symmetry axiom~\cite{shapley1953value}, the Shapley values of agents in the same class (same energy demands) are equal, and hence it is only required to calculate the Shapley values once for each class. \cref{alg:shapley} is used to determine the Shapley values when the community consists of $K$ classes of agents. The algorithm first loops over the number of classes starting from $k=2$ (Line 2, \cref{alg:shapley}). Class 1, the largest of the classes, is skipped for efficiency since it can be computed after the Shapley values of all other classes are determined. Then, for each class, it will iterate through the strata from 0 to $N-1$ (Line 4, \cref{alg:shapley}). Between Line 5 and 12, the Shapley value of the iterating class is updated by adding the marginal contribution. Line 5 of \cref{alg:shapley} shows that it will iterate through every possible subcoalition of the size of the stratum. As shown in \cref{eq:Shapley}, the marginal contribution of an agent in a community is described as $c(\mathcal{S} \cup\{i\})-c(\mathcal{S})$. Hence, it can be seen in Line 5 that the maximum number of agents from the iterating class $k$ is one less than $N_k$, so that the marginal contribution can be computed. Line 6 assures that the formed subcoalition is from stratum $j$.

What makes it possible to compute the Shapley values efficiently for limited number of classes is the \emph{(multivariate) hypergeometric distribution}~\cite{hypergeom172}. The probability mass function of the multivariate hypergeometric distribution $P(\{n_1,..., n_K\}, \{N_1,..., N_K\}, N, n)$ computes the relative frequency of selecting $n_1$ agents from class 1 with the size of $N_1$, $n_2$ agents from class 2 of size of $N_2$, and repeating until class $K$, in a community of $N$ agents. The function is formulated as the following.

\begin{equation}
\label{eq:multivarhyper}
    P(\{n_1,..., n_K\}, \{N_1,..., N_K\}, N, n) = \frac{\prod_1^K \binom{N_k}{n_k}}{\binom{N}{n}}
\end{equation}
where $\sum_1^K N_k = N$ and $\sum_1^K n_k = n$. The hypergeometric distribution allows to compute the probability of certain set of agents to be selected ahead over the chosen agent at the specific stratum. Line 7 in \cref{alg:shapley} shows the step where the probability of the set of agents being ahead at stratum $s$ is computed. (The Shapley value of an agent can then be computed by replacing the relative frequency of a subcoalition in Eq.~\ref{eq:ShapleyStrata}, with the hypergeometric function.)

Line 8 in \cref{alg:shapley} computes the marginal contribution of the agent from class $k$ using the table containing costs of subcoalitions from \cref{alg:CommunityCost}. The marginal contribution in Line 9 is added with the factor of the relative frequency of the subcoalition from stratum $j$. After iterating through every strata and subcoalitions, the value is divided by the total number of strata, which is $N$ (Line 13). It can be seen that computation steps of Lines 8, 9, and 13 are equivalent to~\cref{eq:ShapleyStrata}, with the only difference being the relative frequency is computed using the hypergeometric function.

Line 15 computes the Shapley value of agents in class 1. Since the values of agents of all the other classes are known and the sum of Shapley values of all agents must equal to the community cost, the Shapley value of class 1 is equal to the remaining cost after subtracting the cost distributed to agents in class 2 to $K$ from the community cost, then equally divide it by the agents from class 1, by the efficiency property~\cite{shapley1953value}.

\subsection{Complexity of Shapley value computation} \label{sec:timeComplexity}

\begin{table}[H]
\centering
\resizebox{0.6\linewidth}{!}{%
\begin{tabular}{lcc}
           & \multicolumn{2}{c}{\textbf{Time Complexity}} \\ \toprule
\textbf{Algorithm} & \textbf{Unique} & \textbf{$K$ Classes} \\ \toprule
Exact Shapley                      & $\mathcal{O}(2^N \cdot (N-1))$      & $\mathcal{O}(N^K \cdot (K-1))$*\\ \midrule
Marginal Contribution        & $\mathcal{O}(N)$         & $\mathcal{O}(K)$\\ \midrule
Stratified Expected Values   & $\mathcal{O}(N^2)$       & $\mathcal{O}(K \cdot N)$\\ \midrule
Approx. Shapley RL             & $\mathcal{O}(N \cdot M)$       & $\mathcal{O}(K \cdot M)$\\ \bottomrule
\multicolumn{2}{l}{*\small{Upper bound time complexity}}
\end{tabular}%
}
\caption{Time complexity per algorithm}
\label{tab:runtimeAnalysis}
\end{table}

Table~\ref{tab:runtimeAnalysis} shows the time complexities of exact Shapley values and the three approximation methods used in this study for two scenarios; when the community of size $N$ consists of unique demand profiles and when the number of demand profiles is limited to $K$ classes.

In the case of $N$ unique demands, it was explained previously that it requires $2^N$ steps to compute the Shapley value of an agent. To compute the Shapley values of the whole community, it is required for $N-1$ agents since the value of the last agent can be determined by simply subtracting the sum of the rest of the agents' values from the total cost. This is due to the efficiency property of the Shapley value, in which the sum of the redistributed values equals the total value~\cite{shapley1953value}. Hence, the time complexity of a community of unique agents is $\mathcal{O}(2^N \cdot (N-1))$.  

When the community is restricted to $K$ classes, the number of times the cost function needs to be computed by~\cref{alg:CommunityCost} is equal to the number of all possible combinations of agents which is $(N_1+1) \times ... \times (N_K+1)$ (illustrated in \cref{fig:strata}). Considering w.l.o.g that the classes are ordered by the size, i.e., $N_1 \geq N_2 \geq ... \geq N_K$, and assuming there are at least two non-empty classes, i.e. $K\geq 2$, then it holds that $N_i +1 \leq N,\, \forall i = 1,...,K$. The number of cost function calculations is hence upper bounded by $N^K$. Due to the symmetry property of the Shapley value~\cite{shapley1953value}, it is only required to be computed once per class. Furthermore, it is required to compute $K-1$ times with the same reasoning as in the unique demand profiles scenario. Therefore, it gives the time complexity of $\mathcal{O}(N^K\cdot(K-1))$. While it seems that $\mathcal{O}(N^K\cdot (K-1))$ (for $K$ classes) is large, in fact, for a large $N$ and a small number of classes $K$ this is much smaller than $2^N$, hence in practice, exact Shapley computation with $K$ classes has a much lower computation cost than unique agents.

For the marginal contribution method (\cref{subsec:marginalContribution}), the complexity was determined to be $\mathcal{O}(N)$ for a community of $N$ agents, as it requires to compute the marginal contribution once for every agent. With $K$ classes, however, this is reduced to $\mathcal{O}(K)$. Since agents with the same energy demands are assigned the same cost,~\cref{eq:marginalContribution_new,eq:marginalContribution_normalized} are only required to be computed once for each class.

The time complexity of stratified expected values method (\cref{subsec:SEV}) is $\mathcal{O}(N^2)$ for $N$ unique agents, but it can be reduced to $\mathcal{O}(K \cdot N)$ for $K$ classes. Similarly to the marginal contribution method, it is only required to compute~\cref{eq:stratifiedExpectedValues2} once per class. 

Finally, RL-based Shapley approximation method (\cref{subsec:RL}) has the time complexity reduced from $\mathcal{O}(N \cdot M)$ to $\mathcal{O}(K \cdot M)$, where $M$ is the number of samples per agent chosen by the user, by the same reasoning as the marginal contribution and stratified expected values methods. 
\section{Experimental Comparison} \label{sec:simulation}

For the experimental comparison, the energy demands of 200 households in a realistically-sized energy community in the UK sharing a community wind turbine and battery were used. In this study, experiments are carried out on two scenarios. \cref{subsec:exp1} presents the experimental setup of the first scenario. Here, agents of the community are grouped into two classes based on their annual consumption size (large vs. small energy consumers). The performance of the redistribution methods is tracked with increasing community size, keeping the ratio of large to small consumers constant. \cref{subsec:exp2} presents the experimental setup of the second scenario. Here, agents are grouped into four classes based on their consumption profile throughout a typical day. Again, the performances of the redistribution methods are compared with increasing community size. Finally, \cref{subsec:discussion} provides discussions on the results from the two scenarios. 
All of the experimental code was written in and run with Python 3 (version 3.8.5).

\subsection{Scenario 1: Large and Small Consumers} \label{subsec:exp1}

\begin{figure*}[!t]
    \centering
    \advance\leftskip-1cm
    \advance\rightskip-1cm
    \begin{subfigure}[t]{0.54\textwidth}
        \includegraphics[width=\textwidth]{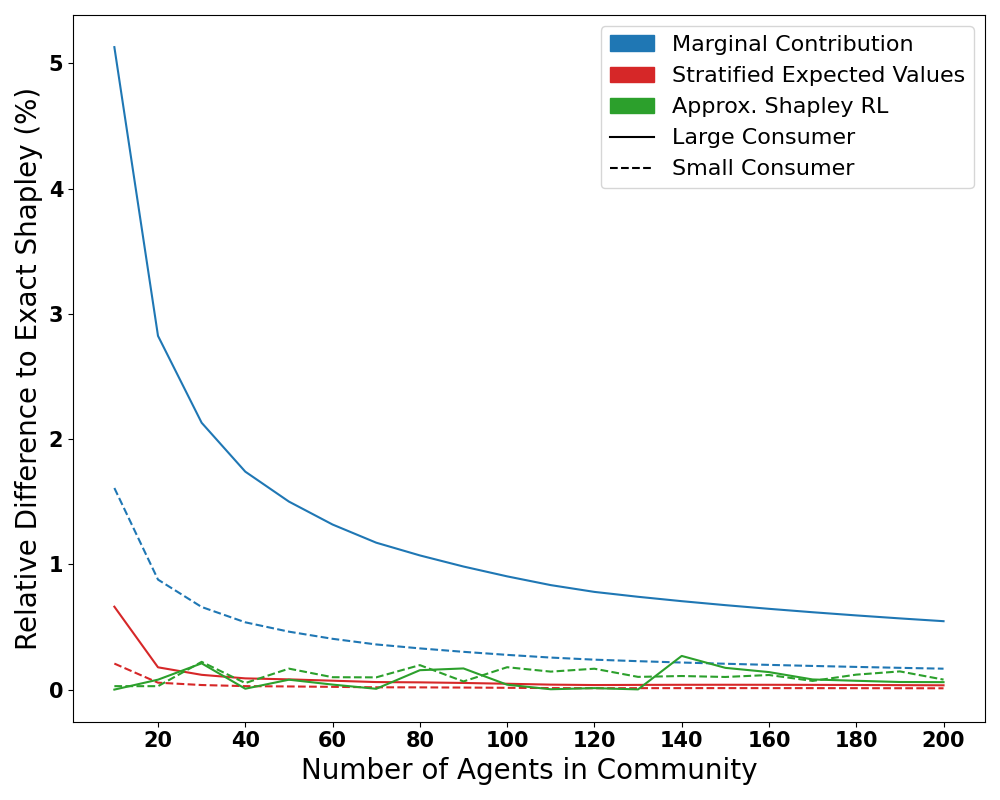}
        \caption{90/10 split}
        \label{fig:90/10split}
    \end{subfigure}
    \begin{subfigure}[t]{0.54\textwidth}
        \includegraphics[width=\textwidth]{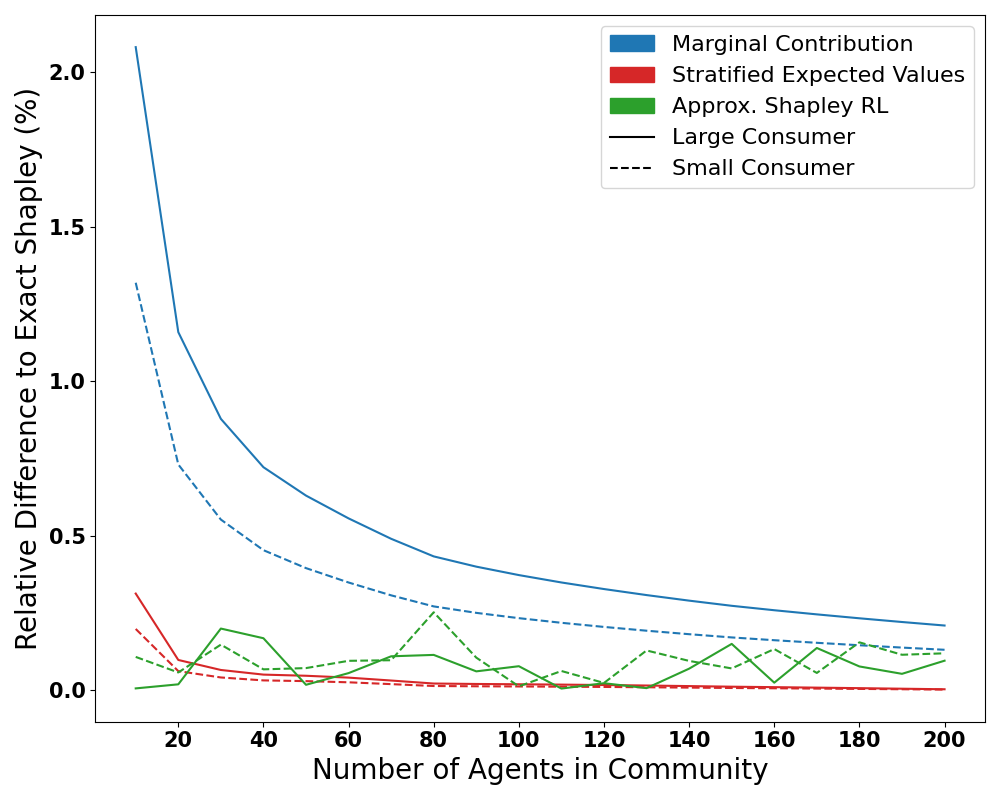}
        \caption{80/20 split}
        \label{fig:80/20split}
    \end{subfigure}
    \caption{Relative differences of the small and large consumer agent profiles to the exact Shapley values for the redistribution methods with increasing size of the community.}
    \label{fig:2class_percentage}
\end{figure*}

\textbf{Dataset and parameters.} 
For the first experimental comparison, the energy demands of 200 households in a realistically-sized energy community in the UK sharing a community wind turbine and battery were used, using the case study from Norbu et al.~\cite{NORBU2021116575} (with the kind permission of the authors). The energy demands of the households are provided for every 30 minutes during one calendar year, which was largely collected in a well-known smart energy demonstrator project in the UK, the Thames Valley Vision project~\cite{TVV}. 
The half-hourly power generated by the wind turbine was calculated based on the power curve of the Enercon E-33 wind turbines~\cite{enercon} and real wind data of the Kirkwall airport weather station in Orkney, Scotland from the UK Met Office Integrated Data Archive System (MIDAS)~\cite{MIDAS_dataset} provided by the British Atmospheric Data Centre (BADC). Furthermore, an import tariff of 16 UK pence/kWh and an export tariff of 0 pence/kWh were used.

\textbf{Demand profiles.} 
Two-hundred prosumers are grouped into two classes of small consumers or large consumers according to the annual energy consumption. Small consumer and large consumer profiles are made from the average half-hourly demands of each group. In this study, two cases are tested; groups split into the 90\% smallest and 10\% largest consumers by total annual demand, respectively, and 80\% smallest and 20\% largest in the second test. The community consists of agents that had small and large consumer profiles, with the corresponding ratios (9:1 or 8:2). 

\textbf{Setup and performance measure.} 
Communities with small and large consumer profiles are used to compare how well the redistribution methods approximate the Shapley values. In the first setting, the ratio of the community is kept to 9:1, and the approximation performances are measured for varying community size of $N=10$ up to $N=200$. Similarly, the ratio is kept constant to 8:2 in the second setting, and performances of varying community size is tested. 

The redistribution methods are compared to the exact Shapley values (the ground truth) of small and large agent profiles. The \emph{relative difference} to the exact Shapley values was used for the comparison. The percentage relative difference of a cost to the Shapley value is defined as the following.
\begin{equation}
\label{eq:relativeDifference}
  RD_{\phi}(\hat\phi_k) = \frac{|\hat\phi_k - \phi_k|}{\phi_k} \times 100
\end{equation}
where $\hat\phi_k$ is the energy cost of agent of class $k$ (in this simulation, either small or large consumer profile) from a particular redistribution method, which are $\overline{MC}_k$, $\overline{SEV}_k$, and $RL_k$. The variable $\phi_k$ is the cost redistributed to class $k$ according to the Shapley value. The relative difference does not only take the magnitude of difference between the approximation method and the exact value, but also considers how large the exact value is. This provides a fairer evaluation between different demand profiles, as demand profiles with naturally large energy cost could have significant approximation error in terms of magnitude only from slight deviation. 

\textbf{Results.}
We investigated whether the size of the community influences how well the redistribution methods approximate the Shapley values. \Cref{fig:2class_percentage} shows the change in relative difference to the exact Shapley values of the redistribution methods with increasing community size up to 200 households while keeping the same ratio of small and large agent profiles. In~\cref{fig:90/10split}, the ratio of small consumer agents and large consumer agents were kept to 9:1, while ~\cref{fig:80/20split} used the ratio of 8:2.

\subsection{Scenario 2: Different Consumption Profiles}\label{subsec:exp2}

\begin{figure*}[t]
    \centering
    \advance\leftskip-0.5cm
    \advance\rightskip-0.5cm
    \includegraphics[width=1.05\linewidth]{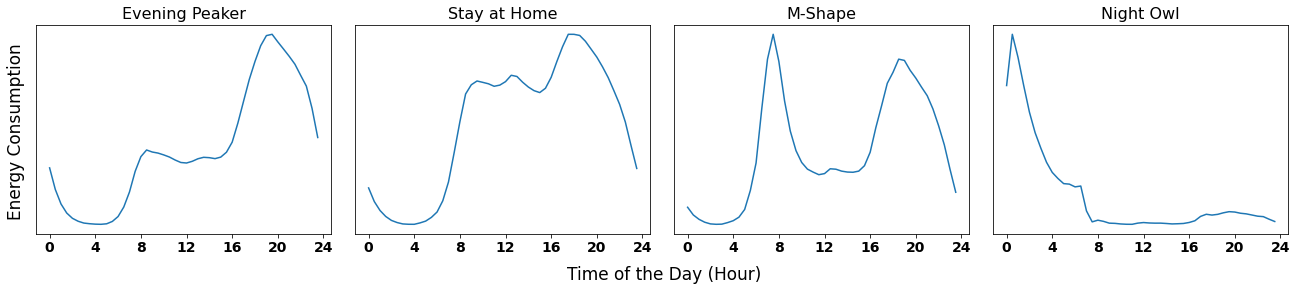}
    \caption{Average daily energy demands of consumer profiles, evening peaker, stay at home, M-shape, and night owl}
    \label{fig:consumptionTypes}
\end{figure*}

\textbf{Dataset and parameters.} 
In the second case study, the demands and the wind data were taken from a dataset in Kaggle\footnote{\url{https://www.kaggle.com/datasets/jeanmidev/smart-meters-in-london}}, a ML data platform. The half-hourly demands of 5567 households in the London area, UK, between November 2011 and February 2014 were recorded by the UK Power Networks during the Low Carbon London project~\cite{lowCarbonLondon}. The corresponding London weather data was provided by Dark Sky~\cite{dark_sky_api_2022}, and the generated power by the wind turbine was calculated using the same method as Norbu et al.~\cite{NORBU2021116575}. 

Both the demands and wind power data were aggregated to generate an averaged half-hourly data points for one year, aligned by the calendar weeks. From the demand data, households were removed if less than 95\% of the data points from the year were missing, resulting in 5251 households left. The remaining miss data points were filled using linear interpolation. 

Similarly to the first case study, an import tariff of 16 UK pence/kWh and an export tariff of 0 pence/kWh were used.

\textbf{Demand profiles.} 
In the second case, the agents were grouped not by their total demands but rather by their consumption profiles over a day (24 hours). Clustering energy consumers into a number of classes, according to their daily consumption profile, is a well-established practice in energy demand modeling~\cite{electronics10030290, 6693793, ZHOU2013103}, used both in research and practice, by energy suppliers. Identifying consumption patterns of customers can help the energy provider to provide customers with recommendation as well as managing energy loads.

The 5251 agents are clustered using K-means clustering from the energy consumption of the winter months. From the resulting clusters, four groups showing distinct behaviours were chosen as consumption classes, and is presented in~\cref{fig:consumptionTypes}.

\cref{fig:consumptionTypes} shows the daily consumption behaviours of the four selected clusters. 
The first cluster on the left shows a increase in demand in the morning, then a large peak in the evening, thus named ``evening peaker''. The second cluster from the left has energy demands increased in the morning and stay high during the day. There is a small evening peak, but has relatively even consumption throughout the day. This group is called ``stay at home'', as it requires certain energy consumption during the day such as heating, computers, and kitchen appliances. The third cluster shows a large peak in the morning followed by a decreased consumption during the day, and a final large peak in the evening. It can be seen that the morning and evening peaks are roughly the same size, and therefore it is named ``M-shaped'' consumers. The fourth cluster had almost no energy demand during the day, but had high demand overnight. Such behaviour is also observed in~\cite{electronics10030290}, a study on clustering consumers on demand profiles. This group was named ``night owl'', and it is more of a rare case, having less than 1\% of the households classified in this study. 

In 2020, due to COVID-19, working from home became the norm, thus it would be of interest to look at a change of behaviour from working at the office to home. Hence looking at consumption behaviours of classes like ``evening peak'' and ``stay at home'' were chosen for this study. Furthermore, to add more variety and create a realistic community, we have included classes with distinct behaviours such as ``M-shape'' and ``night owl'' classes. From grouped agents, the half-hourly energy demands are created for four consumer profiles. The details of clustering and the production of demand profiles are given in~\ref{app:clustering}.

\textbf{Setup and performance measure.} 
Communities with four consumer profiles are used to perform experiments. In this case study, we perform two experiments. In the first experiment, the ratio of the community was kept constant, and the performances of the redistribution methods were tested for community sizes of $N=10$ to $N=200$. We tested two scenarios; one in which the community is concentrated to one class, and one in which the community is more evenly spread out between the classes. In the second setting, the community size is kept constant, but the composition (ratios) of the consumer classes in the community changes.

To compare the performance of the redistribution methods, we used the \emph{average relative difference} to the exact Shapley values. The relative difference to the exact Shapley values is as defined in~\cref{eq:relativeDifference}. The average relative difference to the Shapley value of a redistribution method is the mean relative difference of the community:
\vspace{-0.1cm}
\begin{equation}
\label{eq:averageRelativeDifference}
  RD_{\phi}(\hat\phi) = \frac{1}{N} \sum_k^K N_k \cdot RD_{\phi}(\hat\phi_k) 
\end{equation}
where $\hat\phi$ is a redistribution method with costs $\hat\phi_1,...\hat\phi_K$ assigned to $K$ classes. 

\begin{figure*}[!t]
    \centering
    \advance\leftskip-1cm
    \advance\rightskip-1cm
    \begin{subfigure}[t]{0.54\textwidth}
        \includegraphics[width=\textwidth]{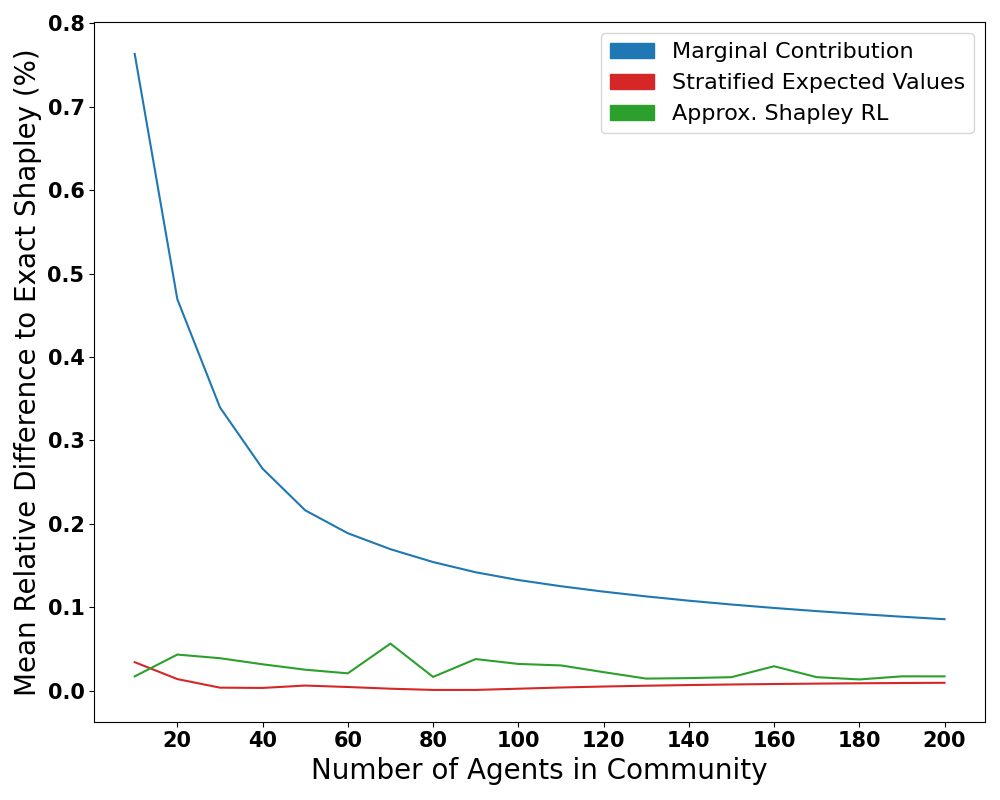}
        \caption{70/10/10/10 split}
        \label{fig:4types-70101010}
    \end{subfigure}
    \begin{subfigure}[t]{0.54\textwidth}
        \includegraphics[width=\textwidth]{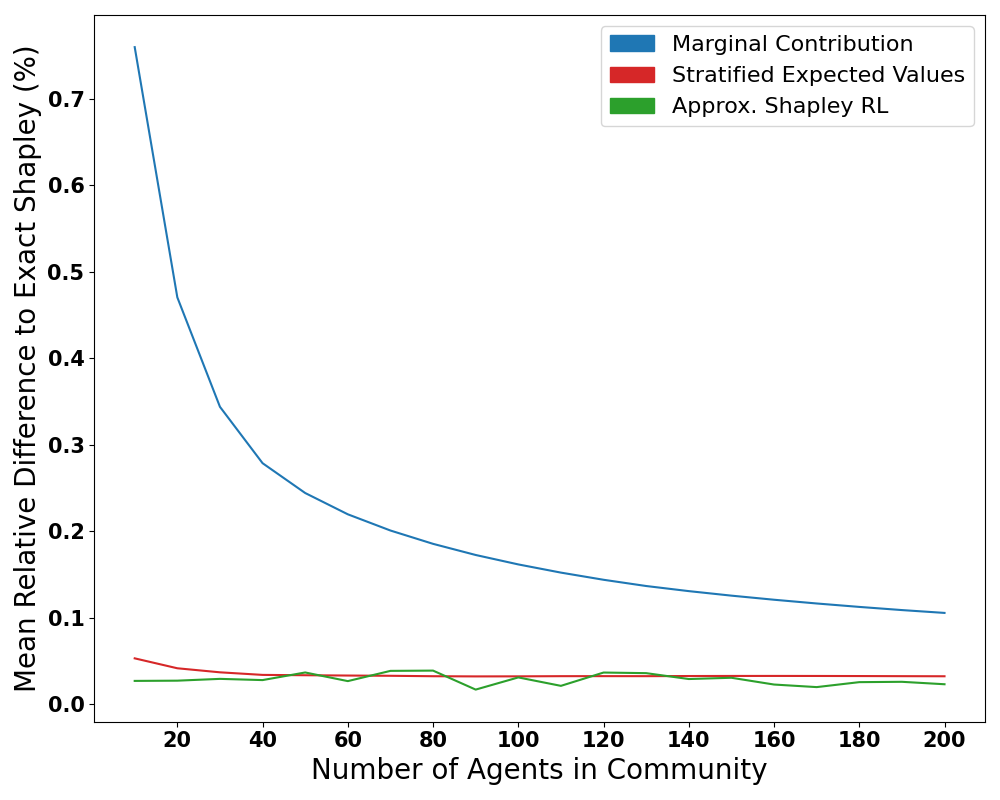}
        \caption{30/30/30/10 split}
        \label{fig:4types-30303010}
    \end{subfigure}
    \caption{Average relative differences to the exact Shapley values for the redistribution methods with increasing community size in a community with four consumption profiles.}
    \label{fig:4types}
\end{figure*}

\begin{figure}[t]
    \centering
    \includegraphics[width=0.5\linewidth]{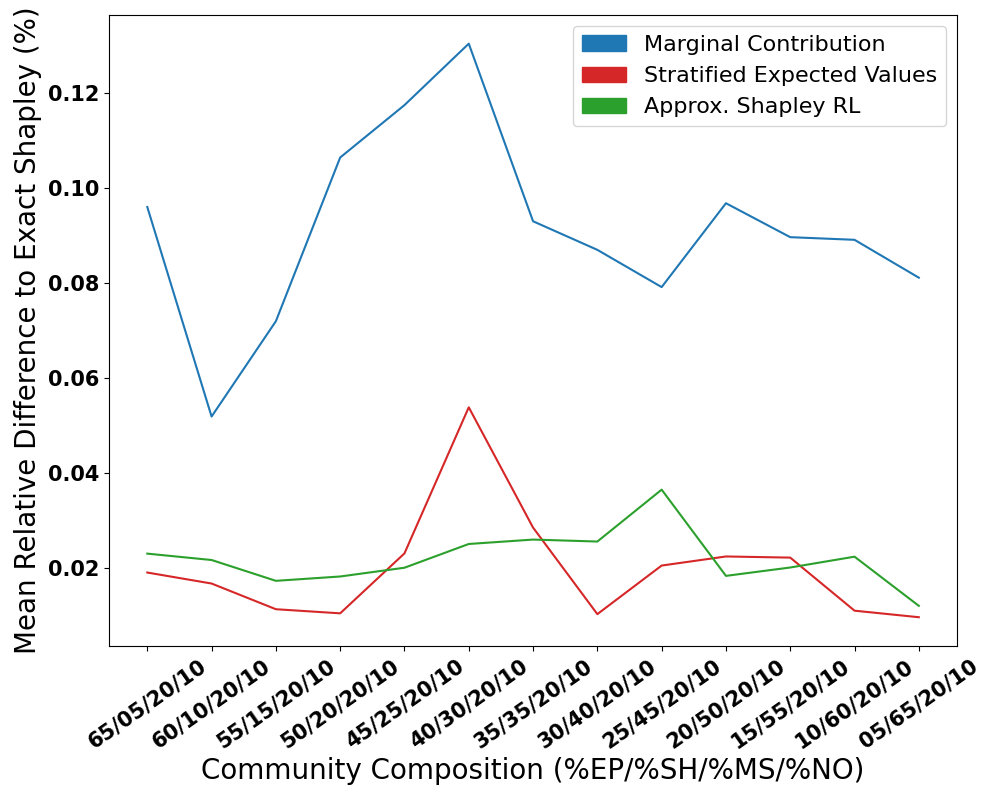}
    \caption{Average relative differences to the exact Shapley values for the redistribution methods of different community compositions with $N=200$.}
    \label{fig:4types-ratio}
\end{figure}

\textbf{Results.} 
\cref{fig:4types} shows the change in average relative differences with increasing community size, starting from $N=10$ up to $N=200$. \cref{fig:4types-70101010} presents the result of the community with compositions of 70\% ``evening peak'', 10\% ``stay at home'', 10\% ``M-shape'', and 10\% ``night owl'', and \cref{fig:4types-30303010} with compositions of 30\% ``evening peak'', 30\% ``stay at home'', 30\% ``M-shape'', and 10\% ``night owl''.
\cref{fig:4types-ratio} shows the change in average relative differences of redistribution methods with change in the composition of the community. The community size was set to 200, and the ratios of ``M-shape'' and ``night owl'' agents were also kept constant to 20\% and 10\% respectively. Initially, the ``evening peak'' class is set to be 65\% of the community and ``stay at home'' class to 5\%. After every run, the ratio of ``evening peak'' is reduced by 5\% and ``stay at home'' increased by 5\%, until the ``stay at home'' makes up 65\% of the community and ``evening peak'' with only 5\%. 
Detailed results showing the breakdown of the performance across the different consumer profiles are presented in~\ref{app:simFigs}.

\subsection{Discussion}
\label{subsec:discussion}
\cref{fig:2class_percentage,fig:4types,fig:4types-ratio} showed that all the tested redistribution methods approximated Shapley values well for a large community. Although the marginal contribution method yielded high errors for small community size (for example, 5\% difference with exact Shapley for large consumer profile in 90/10 split scenario,~\cref{fig:90/10split}), for community size of over 100 prosumers, all methods were below 1\% difference with the exact Shapley in all scenarios. We attribute the smaller difference to the exact Shapley in larger communities to a smoothing effect that a large number of agents have to individual variations. In a large community, the Shapley calculation is dominated by marginal contributions of the agent to already large-sized subcoalitions. Variations in marginal contributions to large subcoalitions are often small, making it possible for less complex methods to approximate well for large communities.

The stratified expected values method outperforms the simpler marginal contribution method in all cases and all scenarios, hence the intuitive hypothesis we formulated in Section~\ref{subsec:SEV} clearly holds. Furthermore, there is a minimal difference in the performances between the stratified expected values and the adaptive sampling methods in most cases. The number of samples per agent was set to 1000 for the adaptive sampling method, meaning that for a case of 100 prosumers community, the adaptive sampling method had a time complexity ten times higher than the stratified expected values method (from~\cref{tab:runtimeAnalysis}). Yet the figures show that the stratified expected values method outperform the adaptive sampling method in many scenarios and perform comparatively overall. 
In fact, paired two-sample t-tests on 2 class experiments with 0.05 significance level showed that the stratified expected values method (90/10 split: $M=0.0339$, $SD=0.0520$. 80/20 split: $M=0.0285$, $SD=0.0471$) had a smaller average relative difference to true Shapley values compared to RL-based adaptive sampling method (90/10 split: $M=0.1110$, $SD=0.0496$. 80/20 split: $M=0.0934$, $SD=0.0444$) for both 90/10 split($t(19) = -3.919$, $p<0.001$) and 80/20 split ($t(19) = -4.236$, $p<0.001$). Furthermore, in the 4 class case with community concentrated to 1 class (\cref{fig:4types-70101010}), the stratified expected values ($M=0.0071$, $SD=0.0070$) outperformed the adaptive sampling method ($M=0.0254$, $SD=0.0114$) ($t(19) = -5.376$, $p<0.001$). Only in the case of evenly spread 4 class community (\cref{fig:4types-30303010}), the adaptive sampling method ($M=0.0283$, $SD=0.0062$) outperformed the stratified expected values ($M=0.0341$, $SD=0.0048$) ($t(19) = 3.173$, $p=0.005$). Hence from these results, it seems that the stratified expected values method does well approximating the Shapley values when the community is concentrated to one class, and outperforms the state-of-the-art sampling method. 

When looking at how the composition of the community affects the performances of the redistribution methods in \cref{fig:4types-ratio}, all three methods approximate Shapley values well (all methods in every scenario less than 0.15\% difference), as the community size is large already. Still, the stratified expected values and the RL-based adaptive sampling methods outperform the marginal contribution method. A paired t-test with 0.05 significance level showed that there is no significant difference between the stratified expected values ($M=0.0199$, $SD=0.0114$) and the adaptive sampling ($M=0.0220$, $SD=0.0056$) methods on their average relative differences to the Shapley values ($t(12) = -0.660$, $p=0.52$). Yet it can be seen from \cref{fig:4types-ratio} that the stratified expected values method has smaller difference to true Shapley values when the community is concentrated on one class, and shows larger errors when the community is more even. This in line with the findings from \cref{fig:4types}. It can be seen in \ref{app:clustering} that most consumers had a ``evening peak'' and made up 60\% of the households studied. Hence it is common to have a community that contains majority of the same consumption behaviour class, making the stratified expected values method desirable in real-world scenarios.

Although it approximates the exact Shapley values very well, a potential disadvantage of the RL-based method (and any method using random sampling) is that the redistributed values can vary every time the algorithm is run. The fluctuating performance of the RL-based method can be seen in~\cref{fig:2class_percentage,fig:4types}. 
In practice, the random output of the method can have an undesirable effect on the \emph{perceived fairness} of the redistribution, as prosumers with the same demand profile can result in being assigned slightly different costs.

\section{Conclusions \& Further Work} \label{sec:conclusion}

While the use of the Shapley value is increasingly popular in energy systems, previous works often sidestep the issue of how it can be efficiently computed in large, realistically-sized settings. The issue is made more pressing by the increasing popularity of community energy projects, where prosumers share joint renewable generation and storage assets and costs. 

This paper aims to close this gap by proposing a new method to efficiently approximate the Shapley value, and characterising both their computational complexity and performance (in terms of distance to the exact Shapley value), using large-scale, realistic case studies of energy communities in the UK. We compare the performance of the new method with an already-existing deterministic method and a non-deterministic, state-of-the-art sampling method. Moreover, in order to develop a ``ground truth'' benchmark to compare these approximations, we propose a novel method to compute the Shapley value \emph{exactly} even for large population sizes by clustering agents into a smaller number of consumption profiles or classes.

Our experimental analysis shows that the relative difference to the true Shapley (while large for a few agents) converges to under 1\% for larger scenarios, basically for all methods considered. In particular, in almost all scenarios studied, the newly proposed stratified expected value method and the state-of-the-art adaptive sampling method perform extremely close to true Shapley values. Interesting to observe that the stratified expected value method performs similarly to the adaptive sampling method~\cite{Obrien_etal_2015} for large populations, although its computational cost is often much lower. In fact, the stratified expected values method outperformed the adaptive sampling method when the community was concentrated to one class, showing a high potential for application in real-world energy communities.

There are a number of directions we find promising to explore in future work. An interesting question to explore is the case when the local distribution network, where the energy community is based is subject to physical capacity constraints (voltage, power)~\cite{norbu_access}. Such constraints could potentially restrict all prosumers to participate in the scheme equally at certain times, and would lead to changes in the coalitional game, as well as in the computation of fair redistribution payments based on the Shapley value. 
Another possible improvement on our current model can be made by providing a more detailed cost calculation of the assets, such as in~\cite{Villa-Arrieta_economicEval}. Although our model takes into account battery degradation for a more accurate annual cost of the battery, a better overall cost estimation can be achieved by considering a longer period of time and taking into account the investment and maintenance costs of these assets, as well as economic factors such as the inflation rate. 

We are also considering extending this work, by implementing our redistribution strategies in a blockchain-enabled smart contract (such as in~\cite{Hua_etal_blockchain,Kirli_etal_blockchain}), which would commit the members of the energy community to a protocol to share the benefits and costs. Based on systematic reviews on blockchains in energy systems~\cite{Andoni_etal_blockchain,Kirli_etal_blockchain}, smart contracts should allow a more decentralised energy system, while preserving the privacy of individual prosumer data, such as demand data. The marginal contribution and stratified expected values methods used in this study already do have favourable characteristics for preserving privacy, as they do not require other prosumers' individual consumption information, only the aggregate consumption of the community. The use of smart contracts could further strengthen the protection of sensitive information and a more secure asset monitoring.

Finally, while this paper focuses on the key topic of Shapley value computation, there are many other fairness concepts that could be explored in energy applications, and it would be relevant to compare their outcome to Shapley values. Conversely, there are many promising concepts proposed in coalitional game theory literature~\cite{Chalkiadakis_al} that - to our knowledge, have been explored much less in energy applications - such as the least-core~\cite{Schulz_least_core_paper} or the nucleolus~\cite{Vesperman_al}. The application and adaptation of such fairness concepts in energy could be a fruitful area, for both research and practice, providing energy communities with the computational tools to make best use of shared energy assets.

\section*{Acknowledgement}

The authors would like to acknowledge the input and contributions of TU Delft master students Daan Hofman, Titus Naber and Kawin Zheng in the initial stages of this work.

In terms of funding, Valentin Robu acknowledges the support of the project ``TESTBED2: Testing and Evaluating Sophisticated information and communication Technologies for enaBling scalablE smart griD Deployment'', funded by the European Union Horizon2020 Marie Skłodowska-Curie Actions (MSCA) [Grant agreement number: 872172]. Sonam Norbu, Merlinda Andoni, Valentin Robu and David Flynn also acknowledge the support of the InnovateUK Responsive Flexibility (ReFLEX) project [ref: 104780]. Merlinda Andoni and David Flynn also acknowledge the support of the UK Engineering and Physical Science Research Council through the National Centre for Energy Systems Integration (CESI) (grant EP/P001173/1) and DecarbonISation PAThways for Cooling and Heating (DISPATCH) project (grant EP/V042955/1).

\bibliographystyle{elsarticle-num-names} 
\bibliography{bibli}

\appendix
\section*{Appendices}
\makenomenclatureB

\nomenclature[B,01]{\(\eta^c\)}{Charging efficiency of battery}
\nomenclature[B,02]{\(\eta^d\)}{Discharging efficiency of battery}
\nomenclature[B,03]{\(\beta \text{ and } \gamma\)}{Parameters for sigmoid function}
\nomenclature[B,07]{\(N^{\text{DoD,max}}_{cycles}\)}{Maximum number of cycles allowed at specific DoD, provided from manufacturer specification}

\nomenclature[C,03]{\(\Delta t\)}{Duration of time period $t$ [hour]}
\nomenclature[C,04]{\(\text{DoD}\)}{Depth of discharge of battery [\%]}
\nomenclature[C,05]{\(\text{DF}^{\text{regular}}\)}{Depreciation factor by regular cycles}
\nomenclature[C,06]{\(\text{DF}^{\text{irregular}}\)}{Depreciation factor by irregular cycles}
\nomenclature[C,08]{\(n^{\text{DoD,regular}}_{cycles}\)}{Number of regular cycles at specific DoD}
\nomenclature[C,09]{\(SoC^{start/end}_{l}\)}{Starting/ending SoC of cycle $l$ [\%]}

\nomenclature[C,10]{\(\pi_{i,j}(m)\)}{Probability of sampling stratum $j$ for agent $i$ at sample $m$}
\nomenclature[C,11]{\(\epsilon(m)\)}{Sigmoid function}
\nomenclature[C,13]{\(\hat \mu_{i,j}\)}{Estimated expected marginal contribution of agent $i$ at stratum $j$ in adaptive sampling}
\nomenclature[C,14]{\(\hat \sigma_{i,j}\)}{Estimated standard deviation of expected marginal contribution of agent $i$ at stratum $j$ in adaptive sampling}
\nomenclature[C,15]{\(h_{i,j}\)}{Count of agent $i$ sampling from stratum $j$ in adaptive sampling}
\nomenclature[C,16]{\(m2_{i,j}\)}{Sum of squared differences from mean of stratum $j$ for agent $i$ in adaptive sampling}
\nomenclature[C,17]{\(mc\)}{Sampled marginal contribution in adaptive sampling}
\nomenclature[C,18]{\(\Delta\)}{Difference between the sampled value and the mean in adaptive sampling}

\nomenclature[C,19]{\(\tilde d_{k}(t)\)}{L2-normalised power demand of class $k$ at time $t$}

\printnomenclatureB
\section{Battery Control Algorithm Equations}
\label{app:batteryControl}
The battery keeps track of its SoC level and the power of the battery, $p^{\text{bat}}(t)$. The power of the battery is negative when charging and positive when discharging. While the battery is charging, the SoC level needs to remain below or equal to the maximum battery capacity, $SoC^{\text{max}}$. In addition, the magnitude of $p^{\text{bat}}(t)$ cannot exceed the maximum (dis)charging power of the battery, $p^{\text{bat, max}}$. These constraints are expressed as the followings. 

\begin{equation}
SoC(t) \leq SoC^{\text{max}}
\label{eq:SoCmax}
\end{equation}
\begin{equation}
\left | p^{\text{bat}}(t) \right | \leq p^{\text{bat, max}}
\label{eq:pbatconstraint1}
\end{equation}

When discharging the battery, similar constraints apply. First, the SoC level cannot go below the minimum battery capacity, $SoC^{\text{min}}$. Second, the magnitude of the battery power may not exceed the maximum discharging power. These constraints are represented as the following.

\begin{equation}
SoC(t) \geq SoC^{\text{min}}
\label{eq:SoCmin}
\end{equation}
\begin{equation}
\left | p^{\text{bat}}(t) \right | \leq p^{\text{bat, max}}
\label{eq:pbatconstraint2}
\end{equation}

The heuristic-based battery control algorithm is described as the following. When the generated power from the RES is greater than the demand ($g(t) > d(t)$), the excess power can be used to charge the battery. However, if the battery is already full or the power exceeds the maximum charging power $p^{\text{bat, max}}$, not all the energy can be stored in the battery, and the surplus power will be sold to the utility grid. The updated $p^{\text{bat}}(t)$, the SoC level, and the exported energy to the utility grid ($e^s(t)$) are determined as the following: 
\begin{equation}
\begin{aligned}
p^{\text{bat}}(t) = -\min(\min((g(t) - d(t)), p^{\text{bat, max}}), \frac{SoC^{\text{max}} - SoC(t-1)}{\eta^c \Delta t})
\end{aligned}
\label{eq:pbatcharge}
\end{equation}
\begin{equation}
SoC(t) = SoC(t-1) - \eta^c p^{\text{bat}}(t) \Delta t
\label{eq:SoCcharge}
\end{equation}
\begin{equation}
e^s(t) = (g(t) - d(t) + p^{\text{bat}}(t)) \Delta t
\label{eq:exportEnergy}
\end{equation}
where $\eta^c$ is the charging efficiency and $\Delta t$ is the duration of time step $t$ in hours. The profit from exporting the energy to the grid can be expressed as the product of the energy exported, $e^s(t)$, and the export tariff, $\tau^{s}(t)$.

Similarly, if the demand is greater than the generated power from RES ($g(t) < d(t)$), the battery is discharged to meet the demand. If the power supplied from discharging the battery is still not enough to meet the demand, energy needs to be imported from the utility grid, denoted as $e^b(t)$. The followings are the updated $p^{\text{bat}}(t)$, SoC levels, and $e^b(t)$ in case of shortage of power:

\begin{equation}
\begin{aligned}
p^{\text{bat}}(t) = \min(\min((d(t) - g(t)), p^{\text{bat, max}}), \frac{\eta^d}{\Delta t} ( SoC(t-1) - SoC^{\text{min}}))
\end{aligned}
\label{eq:pbatdischarge}
\end{equation}
\begin{equation}
SoC(t) = SoC(t-1) - \frac{p^{\text{bat}}(t)}{\eta^d} \Delta t
\label{eq:SoCdischarge}
\end{equation}
\begin{equation}
e^b(t) = (d(t) - g(t) - p^{\text{bat}}(t)) \Delta t
\label{eq:importEnergy}
\end{equation}
where $\eta^d$ is the discharging efficiency. Again, the cost of importing energy from the grid at time $t$ is the product of imported energy, $e^b(t)$, and the import tariff, $\tau^{b}(t)$. 
\section{Battery Degradation Model}
\label{app:batteryDegradation}
In this section, the battery degradation model used in this study and developed by~\citet{NORBU2021116575} is described.

Acceleration of battery degradation caused by frequent charging and discharging operations as well as deep discharging may shorten the battery lifetime to be less than the one specified by the manufacturer. With shortened lifetime, the community needs to replace the battery earlier, incurring additional costs to the households. Hence, the battery degradation model takes this factor into account when calculating the annual cost of the battery for a more accurate representation of the real-world simulation. 

The number of cycles and depth of discharge (DoD) influences battery degradation. A full cycle is defined as SoC returning to the starting value after a discharging and charging phase. On the other hand, a half cycle is defined to be simply the charging or discharging phase. Furthermore, a cycle can be classified as regular or irregular. A regular cycle starts the cycle with the SoC of 100\%, whereas an irregular cycle has the starting SoC to be other than 100\%. Although regular and irregular can have the same DoD (for example, a regular cycle of SoC 100\% discharged to 50\% and charged to 100\%, and an irregular cycle of SoC starting at 80\%, discharged to 30\% and charged to 80\% both have a DoD of 50\%), the battery is depreciated differently. In this study, the rainflow cycle counting algorithm~\cite{DOWNING198231} is used to count the number of full and half cycles as well as identify whether the cycle is regular or irregular.

By counting the number of cycles during the time period, the depreciation factor (DF) of the battery is computed to estimate the battery useful lifetime. As mentioned before, regular and irregular cycles influence the depreciation factor differently. Hence, DF can be defined as the following.

\begin{equation}
    \text{DF} = \text{DF}^{\text{regular}} + \text{DF}^{\text{irregular}}
\end{equation}
where $\text{DF}^{\text{regular}}$ and $\text{DF}^{\text{irregular}}$ are the depreciation factors of regular and irregular cycles respectively. The depreciation factor of regular cycles is determined as follows,

\begin{equation}
    \text{DF}^{\text{regular}} = \sum_{DoD=0\%}^{100\%} \frac{n^{\text{DoD,regular}}_{cycles}}{N^{\text{DoD,max}}_{cycles}}
\end{equation}
where $n^{\text{DoD,regular}}_{cycles}$ is the number of regular cycles at a certain DoD value during the time period, and $N^{\text{DoD,max}}_{cycles}$ is the lifetime of the battery in number of cycles for that DoD value given by the manufacturer. This study used the battery cycle life data of a lithium battery from the work of~\citet{7488267}. The depreciation factor of irregular cycles is expressed as the following.

\begin{equation}
    \text{DF}^{\text{irregular}} = \sum_{l \in L} \text{n}_l \times \left| \frac{1}{N^{\text{DoD}^{eq}(SoC^{Start}_{l}),max}_{cycles}} - \frac{1}{N^{\text{DoD}^{eq}(SoC^{End}_{l}),max}_{cycles}}\right|
\end{equation}
where $L$ is the set of all irregular cycles, and cycle $l$'s starting and ending SoC levels, $SoC^{Start}_{l}$ and $SoC^{End}_{l}$ respectively. The value of $\text{n}_l$ is determined by whether the cycle $l$ is full or half, defined as the following.

\begin{equation}
    \text{n}_l = 
    \begin{cases}
      \frac{1}{2}, & \text{if $l$ is a half cycle}\\
      1, & \text{if $l$ is a full cycle}
    \end{cases}
\end{equation}

Finally, $N^{\text{DoD}^{eq}(SoC^{Start}_{l}),max}_{cycles}$ is the lifetime in number of cycles for $\text{DoD}^{eq}(SoC^{Start}_{l})$, a DoD of a cycle equivalent to starting at 100\% SoC and ending at the value of $SoC^{Start}_{l}$. $\text{DoD}^{eq}(SoC^{Start}_{l})$ is computed as the following.

\begin{equation}
    \text{DoD}^{eq}(SoC^{Start}_{l}) = 100 - \left(\frac{SoC^{Start}_{l}}{SoC^{\text{max}}} \times 100 \right)
\end{equation}

$\text{DoD}^{eq}(SoC^{End}_{l})$ and $N^{\text{DoD}^{eq}(SoC^{End}_{l}),max}_{cycles}$ are similarly computed. 

DF resulting from the former calculation is used in~\cref{eq:batteryCost} for computing the cost of battery during the time period. 
\section{Full Algorithm of Adaptive Sampling Shapley Approximationx} \label{app:RL_approx}

In this section, the details of the RL-based Shapley approximation algorithm by~\citet{Obrien_etal_2015} from~\cref{subsec:RL} are described. 

Prior to running the algorithm, the number of samples per agent, $M$, is predefined by the user. Then, for each agent $i$, the estimated expected marginal contributions of stratum $j$, $\hat \mu_{i,j}$ is initialized to 0. Similarly, the count of how many times stratum $j$ was visited, $h_{i,j}$, and the sum of squared differences from the current mean of stratum $j$, $m2_{i,j}$, are initialized to 0. Finally, the estimated standard deviation of the marginal contributions of stratum $j$, $\hat \sigma_{i,j}$ is initialized to a very large value (here, 10000). 

For every sample of agent $i$, the stratum $j$ is selected according to the probabilities of each stratum at the certain sample. The probability of stratum $j$ being selected for agent $i$'s $m$-th sample $\pi_{i,j}(m)$ is given as the following.
\begin{equation}
    \pi_{i,j}(m) = \frac{\epsilon(m)}{N} + (1 - \epsilon(m))\frac{\hat \sigma_{i,j}}{\sum_{s=0}^{N-1} \hat \sigma_{i,s}}
\label{eq:stratumPorbability}
\end{equation}
where ${\epsilon(m)}$ is a double sigmoid function which helps exploration at the beginning (small $m$ value) and exploitation near the end (large $m$), defined as the following.
\begin{equation}
    \epsilon(m) = 1 + \frac{1}{1+e^{\frac{\gamma}{\beta}}} - \frac{1}{1+e^{- \frac{m - \gamma M}{\beta M}}}
\label{eq:epsilon}
\end{equation}
The parameters $\beta$ and $\gamma$ were set to 0.075 and 0.2 respectively during the experiment since it was found that the sigmoid function with these parameter setting to approximate ideal sampling well~\cite{Obrien_etal_2015}. To approximate the Shapley value well, more samples may be required from certain strata in which the marginal contribution values can vary highly. On the other hand, if the marginal contributions are similar within a stratum, such strata would not need large samples to be approximated.~\cref{eq:stratumPorbability} helps to distribute the samples in such a way.

Once a stratum is chosen, a subcoalition $\mathcal{S}$ is chosen randomly from the selected stratum (i.e., $|\mathcal{S}|=j$), and the marginal contribution of agent $i$ to the subcoalition is computed as $mc = c(\mathcal{S} \cup \{i\}) - c(\mathcal{S})$. The difference between the sampled marginal contribution and the estimated expected marginal contribution, $\Delta = mc - \hat \mu_{i,j}$ is also calculated. Then, the variables are updated after each sample as followings.
\begin{equation}
    h_{i,j} \leftarrow h_{i,j}+1
\end{equation}
\begin{equation}
    \hat \mu_{i,j} \leftarrow \hat \mu_{i,j} + \frac{\Delta}{h_{i,j}}
\end{equation}
\begin{equation}
    m2_{i,j} \leftarrow m2_{i,j} + \Delta (mc - \hat \mu_{i,j})
\end{equation}
Furthermore, $\hat \sigma_{i,j}$ is also updated if the stratum has been visited more than once by agent $i$, as the following.
\begin{equation}
    \hat \sigma_{i,j} \leftarrow \sqrt{\frac{m2_{i,j}}{h_{i,j} - 1}}
\end{equation}

Once all the variables are updated, it moves on to the next sample. After $M$ samples are taken for agent $i$, the redistributed energy cost according to RL-based Shapley Approximation method, $RL_i$ is calculated by taking the mean of expected marginal contributions over the strata, i.e.,
\begin{equation}
    RL_i = \frac{1}{N}\sum^{N-1}_{j=0} \hat \mu_{i,j}
\end{equation}

A difference from the original work is that strata 0 and $N-1$ are only chosen once each since no matter how many times these strata are sampled, they will always have the same marginal contributions (since there is only 1 possible subcoalition). By doing so, more samples can be used in different strata, allowing the algorithm to make use of samples slightly more efficiently.

\section{Clustering \& Consumer Profiles} \label{app:clustering}

For the experiments from~\cref{subsec:exp2}, the 5251 consumers from the Kaggle dataset are clustered with the following steps. The half-hourly energy demands were normalised using L2 normalisation for each agent. From the normalised dataset, only the winter months of the UK were kept, which were January, February, November, and December. This is because the energy consumption is larger during winter, and hence clearer consumption patterns should be observable. From the remaining data, the days on and near Christmas and new year were also removed, which were January 1st to 6th and December 22nd to 31st. This is because many households would likely have abnormal consumption behaviour, such as being absent from home during these days or not working. Finally, only weekdays (from Monday to Thursday) were kept so that working days are what is being clustered. The remaining data consisted of $60 \text{ days} \times 48 \text{ normalised demands per day} = 2880 \text{ datapoints}$ per agent. The new dataset is aggregated so that it contains averaged half-hourly normalised demands of each agent (48 datapoints per agent). Then, K-means clustering is used to group the agents on winter days' energy consumption. The elbow method was used to determine the number of clusters to be 9.

\begin{figure}[H]
    \centering
    \includegraphics[width=0.97\linewidth]{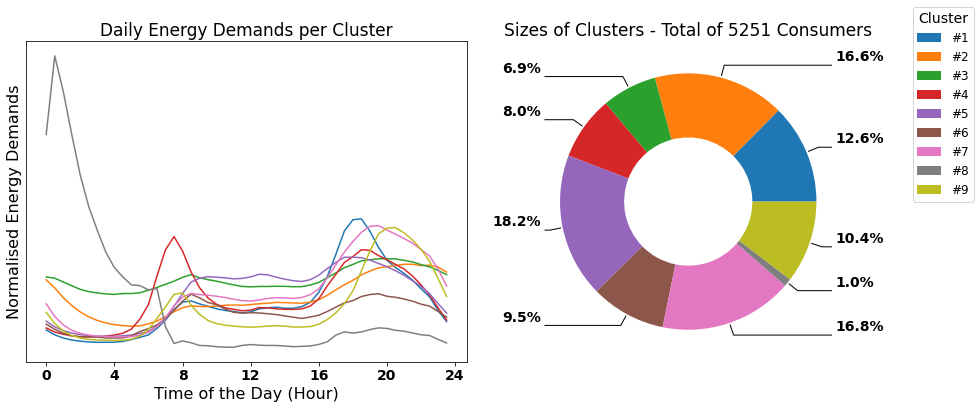}
    \caption{Daily energy demands and the relative sizes of the 9 consumption clusters.}
    \label{fig:clustering_details}
\end{figure}

The average daily demands and the sizes of the 9 resulting clusters are presented in~\cref{fig:clustering_details}. It can be seen from the figure that although the detailed consumption behaviours are unique, many classes shows similarity in terms of having a small morning peak in the morning and a evening peak. The cluster classes 1, 2, 3, 7, and 9 shows variations of such behaviour, and it makes up more than 60\% of the 5251 consumers. Note that the data is before the onset of COVID-19 pandemic, and therefore working from the office during the day was common, making the consumption during the day low. We have chosen the ``evening peak'' class and used the demand profile of class 7 as the representative as it is the largest class out of the evening peaking classes. Next, class 5 shows a high and constant consumption during the day. This behaviour can be thought of as the household working at home as staying at home requires certain energy consumption during the day such as for heating, computers, and kitchen appliances, resulting in an overall high consumption that is not too concentrated in the evening. We suspect that consumers with such consumption behaviour has increased since the outbreak of COVID-19, and hence class 5 was chosen as the ``stay at home'' class to see the impact of such a behaviour in the community. Classes 4 and 6 are unique from the rest of the classes as they both have two almost equal peaks in the morning and in the evening, though the morning peak is slightly larger. This behaviour is a minority in the community, yet 17.5\% of the consumers belong in these groups. Hence, these consumers needs to be represented in the community as well, and the demand profile of class 4 is used for the ``M-shaped'' class in this paper as it shows the behavior more clearly. Finally, one clear outlier cluster is class 8. Although only 1\% of the consumers belongs to this class, it has the most distinct behaviour from the rest of the classes. This class has a very high demand during the night, yet very little demand during the day. This ``night owl'' class was chosen as the fourth class for this study as the outliers of the community. 

Once the clustering is done and consumption behaviours are identified, half-hourly energy demands for 1 year for the four consumer profiles need to be identified. For each class, the L2 normalised half-hourly demands of the year (no days removed and not aggregated) were averaged over the agents belonging in that cluster. 
The yearly demand of the community of these agents is kept equal to $N$ times the average yearly consumption of the 5251 agents from the original dataset. Given the sizes of the four classes $N_{ep}$, $N_{sh}$, $N_{ms}$, and $N_{no}$ and $N = N_{ep}+N_{sh}+N_{ms}+N_{no}$, the restriction is represented as the following.

\begin{equation}
\frac{N}{200} \sum_{i = 1}^{200} \sum_{t=1}^{T} d_{i}(t) = \sum_{k \in \{ep, sh, ms, no\}} \sum_{t=1}^{T} N_{k} d_{k}(t)
\end{equation}

From the restriction, the demand profiles of each class can be computed as the following, given $\tilde d_k(t)$, the normalised 
demand of class $k$ at time step $t$.

\begin{equation}
  \begin{array}{l}
    \displaystyle d_k(t) = \tilde d_k(t) \cdot \frac{\frac{N}{200} \sum_{i = 1}^{200} \sum_{t=1}^{T} d_{i}(t)}{\sum_{q \in \{ep, sh, ms, no\}} \sum_{t=1}^{T} N_{q} \tilde d_q(t)},\quad
    \displaystyle \forall k \in \{ep,sh, ms, no\}, \forall t \in \{1, ..., T\}
  \end{array}
\end{equation}

\section{Additional experimental results}\label{app:simFigs}

Additional experimental results are presented here, in particular in terms of the performance of the Shapley approximations for the different types of consumer classes in ~\cref{subsec:exp2}.

First, \Cref{tab:Values2class} shows the comparison of redistributed yearly energy costs of a small and large consumer agent for both 90/10 and 80/20 splits of 200 households community from~\cref{subsec:exp1}. It can be seen that, for both splits, the energy costs of the small consumer agent are very close across the four redistribution methods, although the redistributed costs based on the marginal contribution are slightly higher than the other methods. On the other hand, the marginal contribution method is lower than the other methods for the large consumer agent. The values for the large consumer agent are still similar across the methods, but the difference between the marginal contribution methods and the Shapley value is more significant compared to the small consumer agent. 

Similarly,~\Cref{tab:Values4class} shows the comparison of redistributed costs of four consumption profiles in a community of 200 agents from~\cref{subsec:exp2}. All methods approximate the Shapley value well for every consumption profile. Especially, the differences between the Shapley values for ``evening peaker'', ``stay at home'', and``M-shape'' agents are very small. For the ``night owl'' agent, the difference is slightly larger for marginal contribution method, but the difference is still only about 0.6\% for the 70/10/10/10 composition and 0.5\% for the 30/30/30/10 composition. 

\begin{table}[H]
\renewcommand{\arraystretch}{1.5}
\centering
\aboverulesep = 0pt
\belowrulesep = 0pt
\caption{Redistributed costs of small and large consumer agents in a community of 200 agents.}
\label{tab:Values2class}
\begin{tabular}{llc|cc}
\toprule
\textbf{Split} & \multicolumn{2}{c|}{\textbf{90 / 10}}  & \multicolumn{2}{c}{\textbf{80 / 20}}  \\ \toprule
 \textbf{Agent Class} & \textbf{Small} & \textbf{Large}  & \textbf{Small} & \textbf{Large}  \\ \toprule
  Shapley      & 245.30           & 674.88  & 221.78          & 554.18    \\ \midrule
  Marginal Contribution      & 245.71           & 671.19   & 222.07          & 553.02        \\ \midrule
  Stratified Expected Values & 245.33             & 674.64   & 221.78             & 554.16      \\ \midrule
 Approx. Shapley RL         & 245.50             & 674.47  & 222.04            & 553.65\\ \toprule  
\end{tabular}
\end{table}

\begin{table}[H]
\renewcommand{\arraystretch}{1.5}
\centering
\aboverulesep = 0pt
\belowrulesep = 0pt
\caption{Redistributed costs of four consumer profiles (``evening peak'':EP, ``stay at home'':SH, ``M-shaped'':MS, and ``night owl'':NO) in a community of 200 agents with two different compositions.}
\label{tab:Values4class}
\begin{tabular}{llccc|cccc}
\toprule
\textbf{Split} & \multicolumn{4}{c|}{\textbf{70 / 10/ 10 / 10}}  & \multicolumn{4}{c}{\textbf{30 / 30 / 30 / 10}}  \\ \toprule
 \textbf{Agent Class} & \textbf{EP} & \textbf{SH}  & \textbf{MS} & \textbf{NO}  & \textbf{EP} & \textbf{SH}  & \textbf{MS} & \textbf{NO}  \\ \toprule
  Shapley      & 341.55           & 369.41  & 325.76          & 239.86  & 334.90           & 364.49  & 321.34          & 234.65    \\ \midrule
  Marginal Contribution      & 341.70           & 369.45  & 325.84          & 238.65  & 335.04           & 364.71  & 321.47          & 233.18    \\ \midrule
  Stratified Expected Values & 341.56           & 369.41  & 325.77          & 239.72  & 334.92           & 364.61  & 321.35          & 234.19    \\ \midrule
 Approx. Shapley RL         & 341.59           & 369.48  & 325.81          & 239.73  & 334.96           & 364.62  & 321.40          & 234.62    \\ \toprule  
\end{tabular}
\end{table}

\begin{figure}[H]
\centering
    \begin{subfigure}[c]{0.49\linewidth}
        \includegraphics[width=\linewidth]{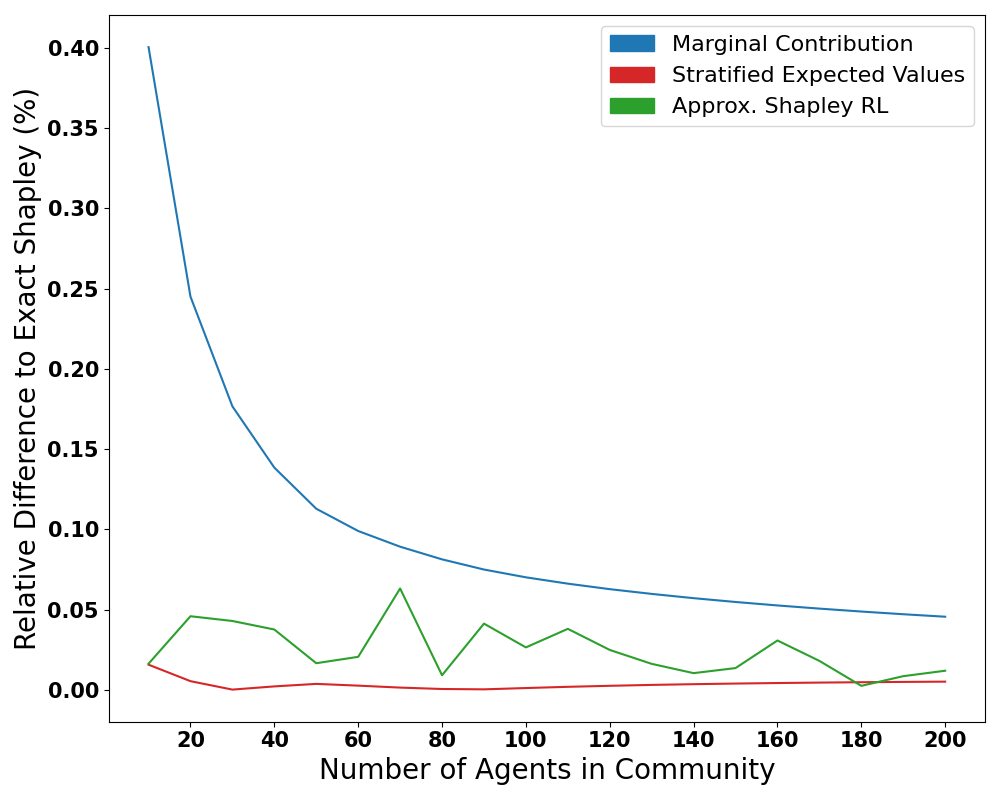} 
        \caption{Evening Peak}
        \label{fig:70101010a}
    \end{subfigure}\hfill    
        \begin{subfigure}[c]{0.49\linewidth}
        \includegraphics[width=\linewidth]{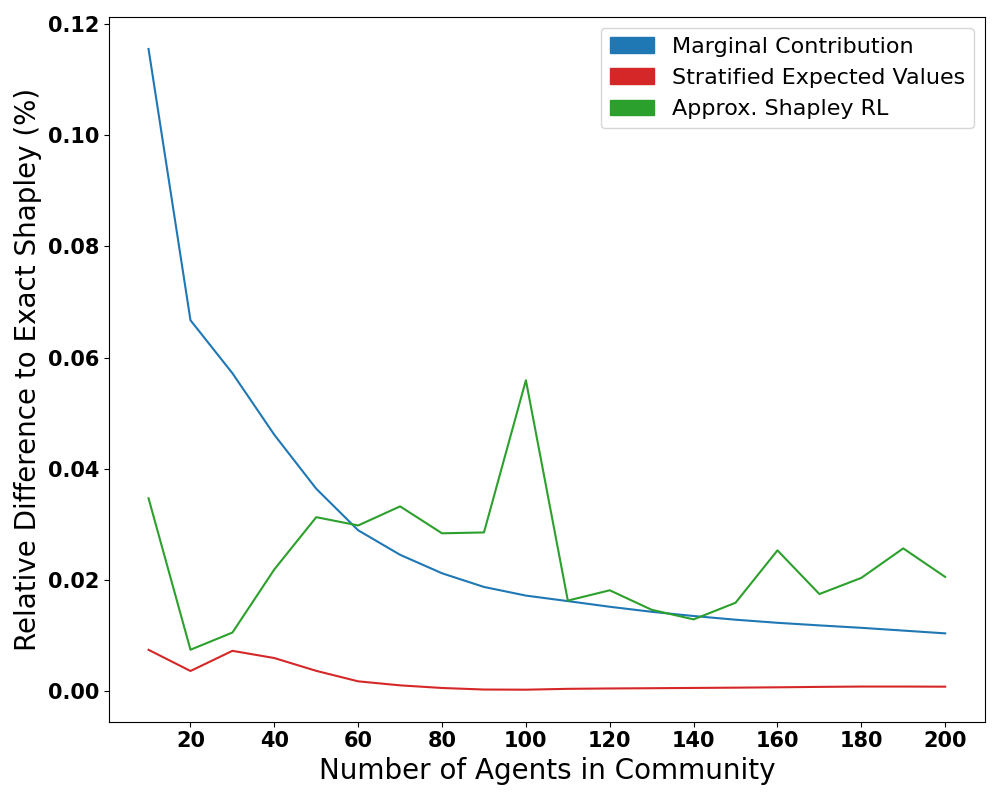}
        \caption{Stay at Home}
        \label{fig:70101010b}
    \end{subfigure}
        \begin{subfigure}[c]{0.49\linewidth}
        \includegraphics[width=\linewidth]{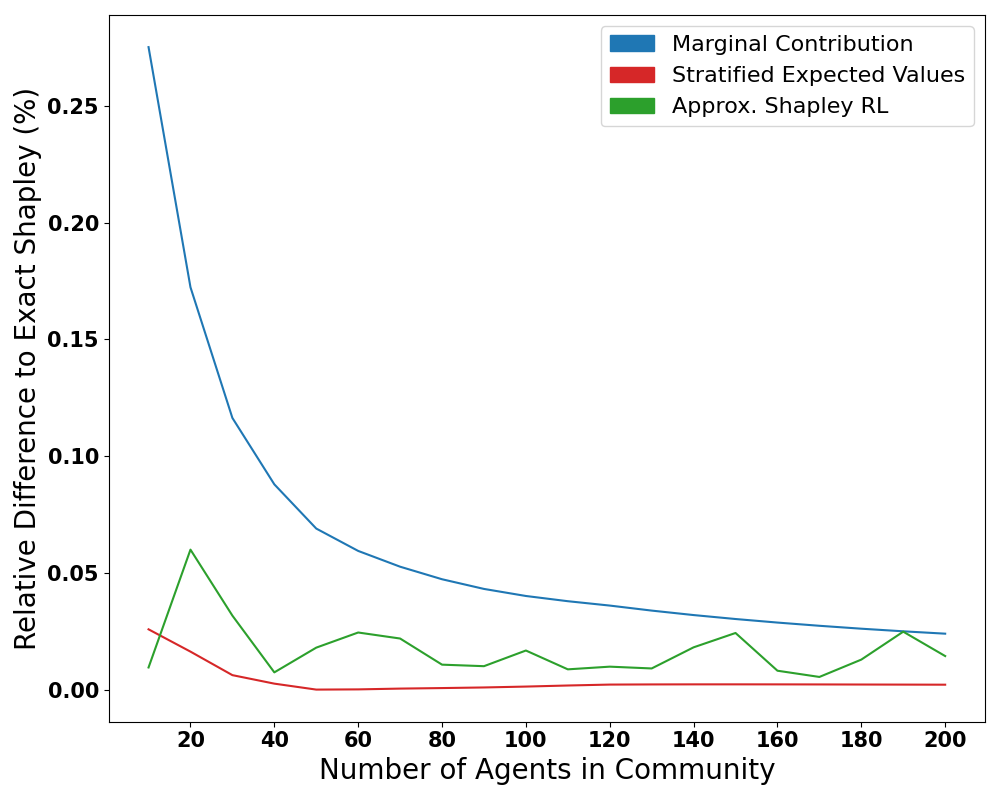}
        \caption{M-shape}
        \label{fig:70101010c}
    \end{subfigure}\hfill    
        \begin{subfigure}[c]{0.49\linewidth}
        \includegraphics[width=\linewidth]{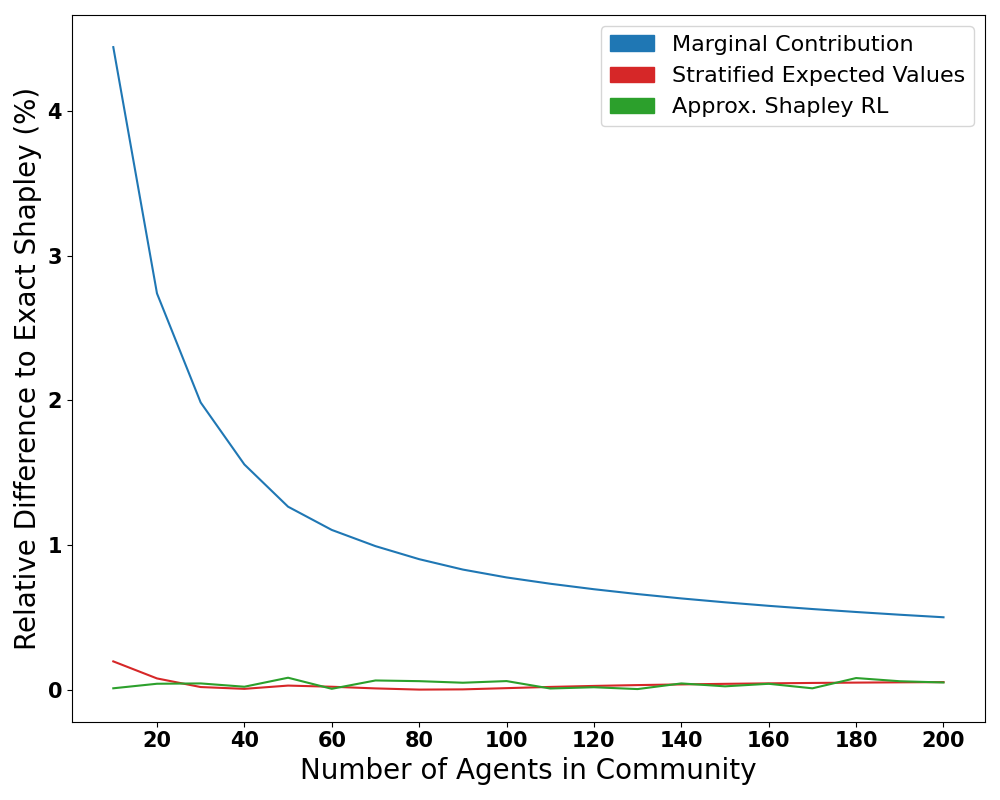}
        \caption{Night Owl}
        \label{fig:70101010d}
    \end{subfigure}
    \caption{Individual relative differences to the exact Shapley values for the redistribution methods of four consumer profiles (70\% ``evening peak'', 10\% ``stay at home'', 10\% ``M-shape'', and 10\% ``night owl'') with increasing size of the community.}
    \label{fig:70101010-eachType}
\end{figure}

\cref{fig:70101010-eachType} contains the relative differences to the exact Shapley values of the redistribution methods for each consumer profile (``evening peak'', ``stay at home'', ``M-shaped'', and ``night owl'') with increasing community size and the composition of 70/10/10/10 (concentrated community), from~\cref{subsec:exp2}. It can be seen that for all classes, the marginal contribution method has similar performance curve, where the error is very large for small community size but decreases quickly as the community size increases. Still, the error value is significantly larger for ``night owl'' class (\cref{fig:70101010d}) than the other three. However, it is also noticeable for ``stay at home'' agent (\cref{fig:70101010b}) that it outperforms the state-of-the-art adaptive sampling method from medium-sized communities ($N \geq 70$). Stratified expected values method, on the other hand, shows high similarity to the exact Shapley values for any community size, and outperforms the simpler marginal contribution method in every scenario. Furthermore, it also outperforms the computationally larger adaptive sampling method in most scenarios, especially for ``evening peak'' (\cref{fig:70101010a}), ``stay at home'' (\cref{fig:70101010b}), and ``M-shape'' (\cref{fig:70101010c}) agents. The RL-based adaptive sampling method also showed high performance regardless of the community size. Yet, it can be seen from all classes that the performance can significantly vary between runs or scenarios due to the random nature of the method. 

\begin{figure}[H]
\centering
    \begin{subfigure}[c]{0.49\linewidth}
        \includegraphics[width=\linewidth]{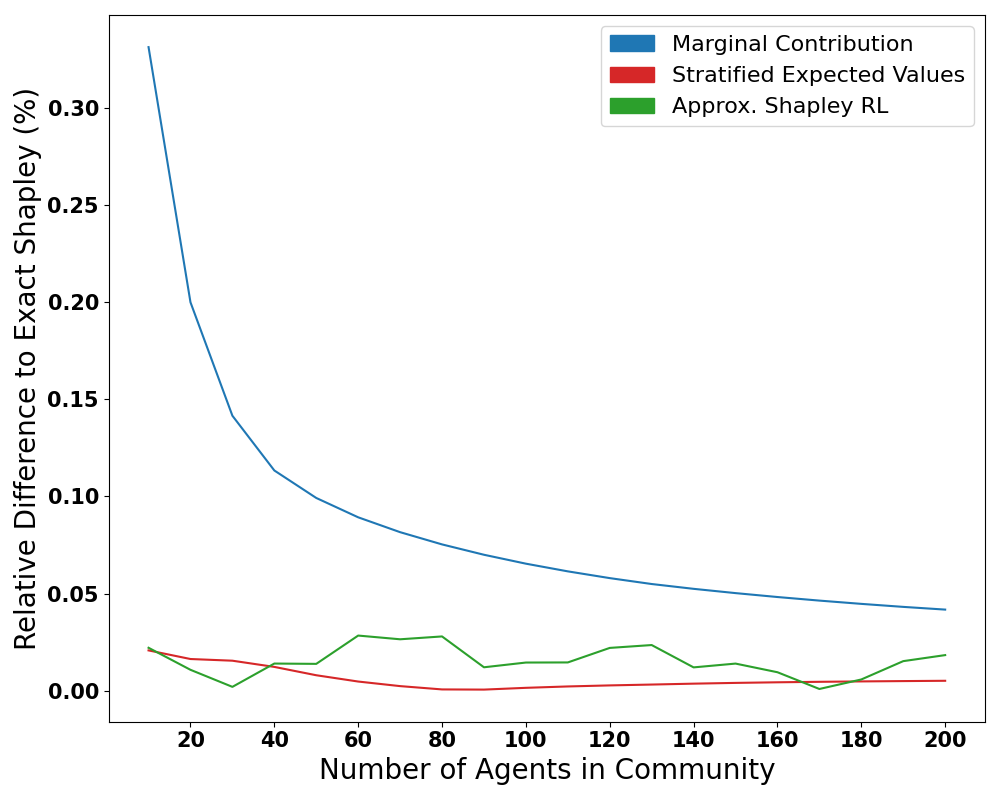} 
        \caption{Evening Peak}
        \label{fig:30303010a}
    \end{subfigure}\hfill    
        \begin{subfigure}[c]{0.49\linewidth}
        \includegraphics[width=\linewidth]{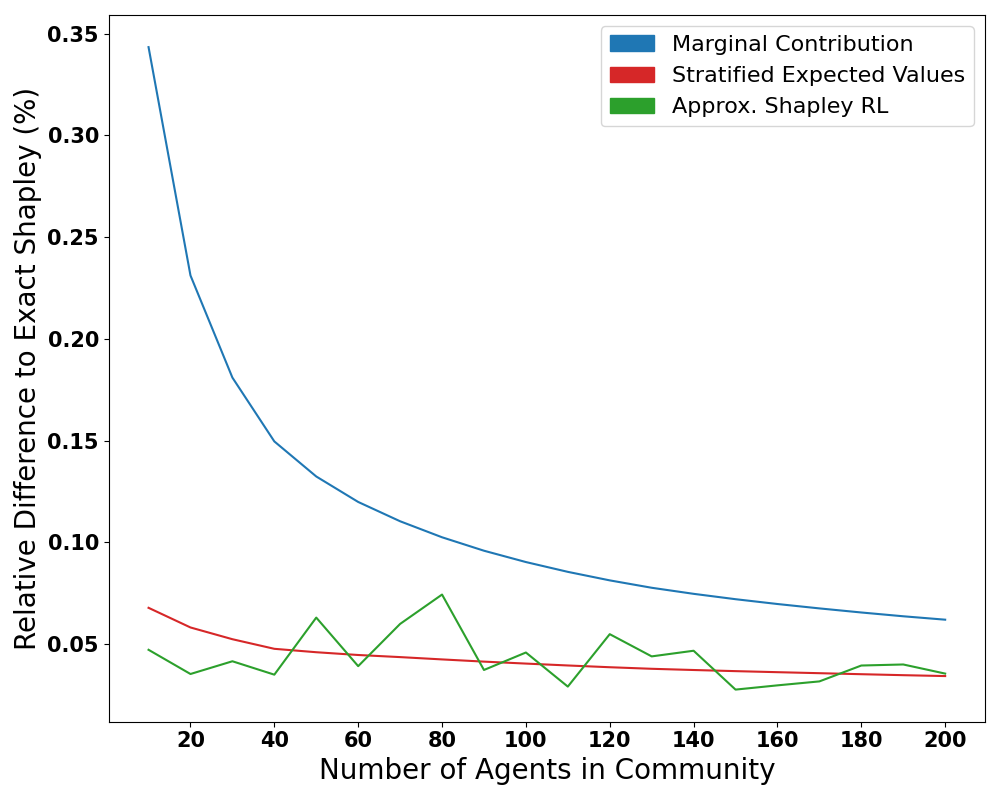}
        \caption{Stay at Home}
        \label{fig:30303010b}
    \end{subfigure}
        \begin{subfigure}[c]{0.49\linewidth}
        \includegraphics[width=\linewidth]{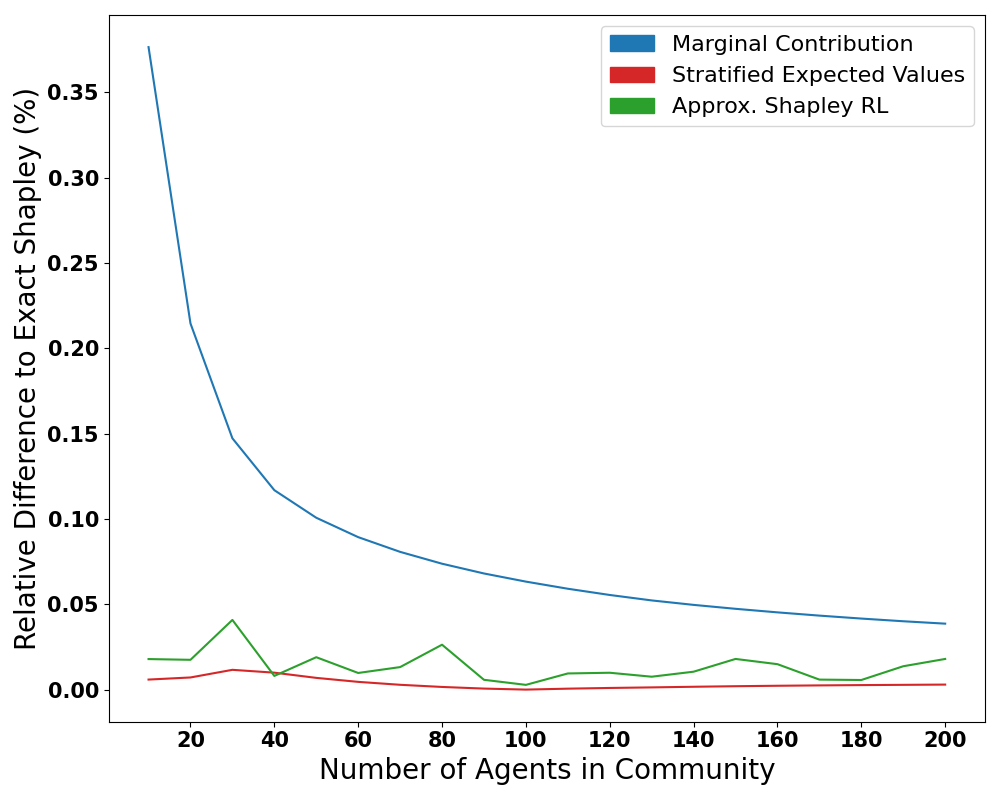}
        \caption{M-shape}
        \label{fig:30303010c}
    \end{subfigure}\hfill    
        \begin{subfigure}[c]{0.49\linewidth}
        \includegraphics[width=\linewidth]{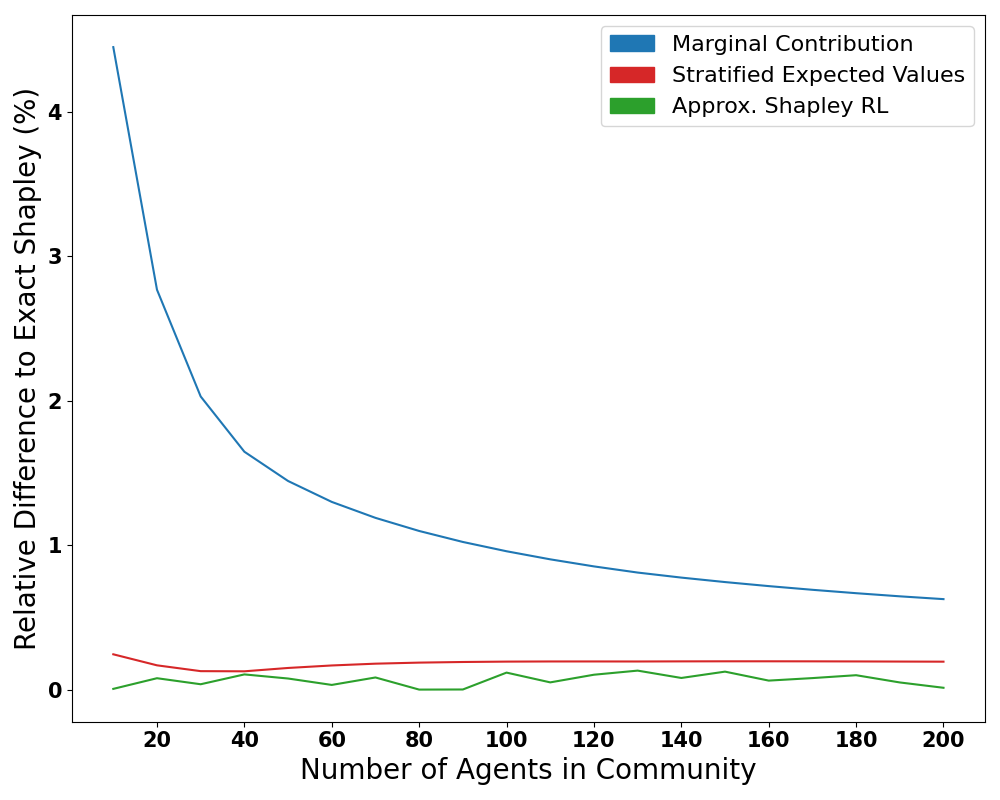}
        \caption{Night Owl}
        \label{fig:30303010d}
    \end{subfigure}
    \caption{Individual relative differences to the exact Shapley values for the redistribution methods of four consumer profiles (30\% ``evening peak'', 30\% ``stay at home'', 30\% ``M-shape'', and 10\% ``night owl'') with increasing size of the community.}
    \label{fig:30303010-eachType}
\end{figure}

Similarly, \cref{fig:30303010-eachType} shows the relative differences of each class with increasing community size and community composition of 30/30/30/10 (even community) from \cref{subsec:exp2}. Again, the marginal contribution method shows fast improvement in performances as the community size grow for every class. Still, it is outperformed by the stratified expected values and the adaptive sampling methods. The stratified expected values does not perform as well as in \cref{fig:70101010-eachType}, only outperforming the adaptive sampling method on ``M-shape'' class (\cref{fig:30303010c}). Although there seems little difference in the overall performances between the two methods for ``evening peak'' (\cref{fig:30303010a}) and ``stay at home'' (\cref{fig:30303010b}) classes, the adaptive sampling outperforms the stratified expected values for ``night owl'' class (\cref{fig:30303010d}) with relatively larger values. In fact, the ``night owl'' class is the main contributor of the overall error of the stratified expected values method in \cref{fig:4types-30303010}.

\begin{figure}[H]
\centering
    \begin{subfigure}[c]{0.49\linewidth}
        \includegraphics[width=\linewidth]{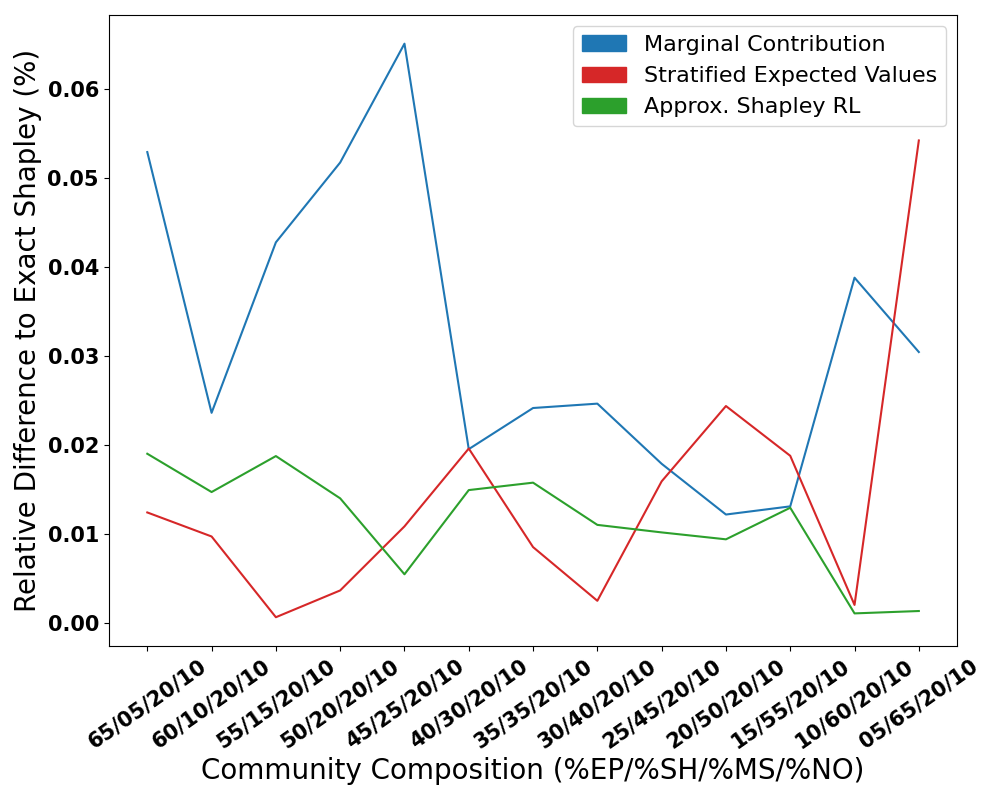} 
        \caption{Evening Peak}
        \label{fig:4types-ratio-eachTypea}
    \end{subfigure}\hfill    
        \begin{subfigure}[c]{0.49\linewidth}
        \includegraphics[width=\linewidth]{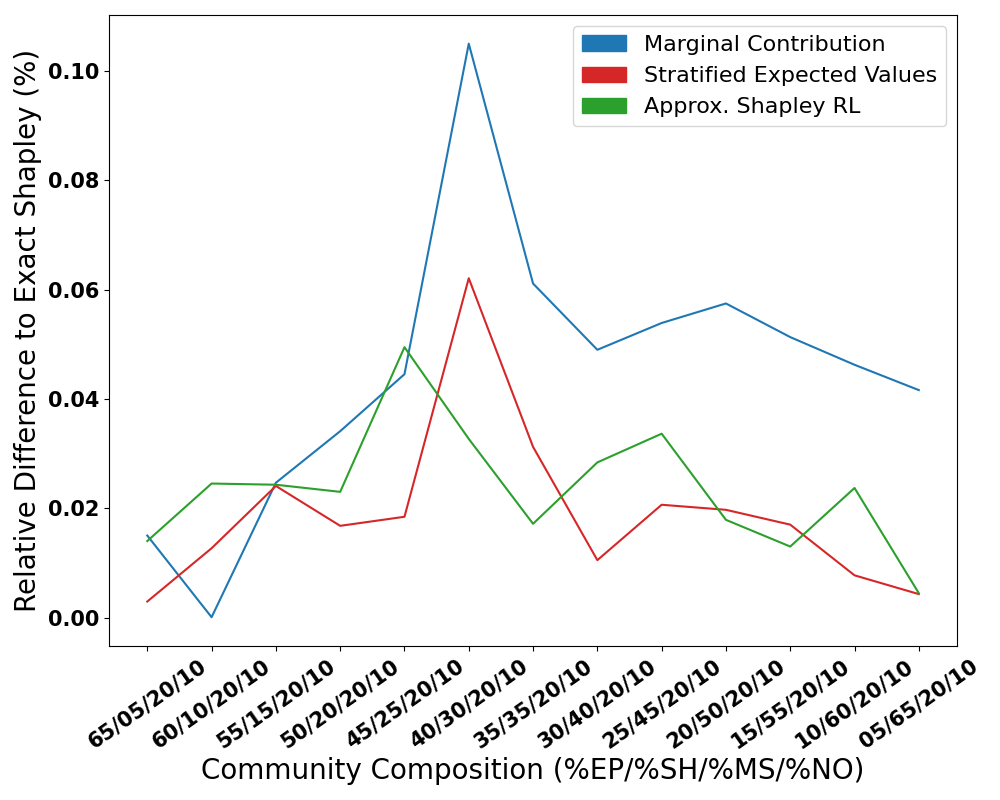}
        \caption{Stay at Home}
        \label{fig:4types-ratio-eachTypeb}
    \end{subfigure}
        \begin{subfigure}[c]{0.49\linewidth}
        \includegraphics[width=\linewidth]{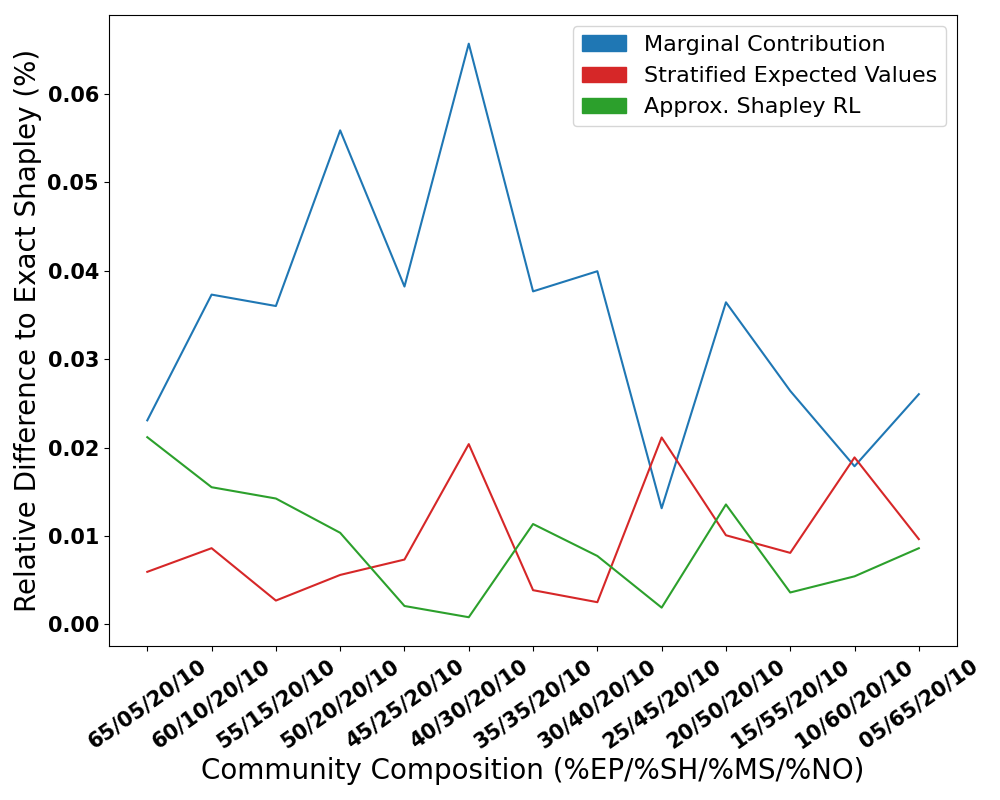}
        \caption{M-shape}
        \label{fig:4types-ratio-eachTypec}
    \end{subfigure}\hfill    
        \begin{subfigure}[c]{0.49\linewidth}
        \includegraphics[width=\linewidth]{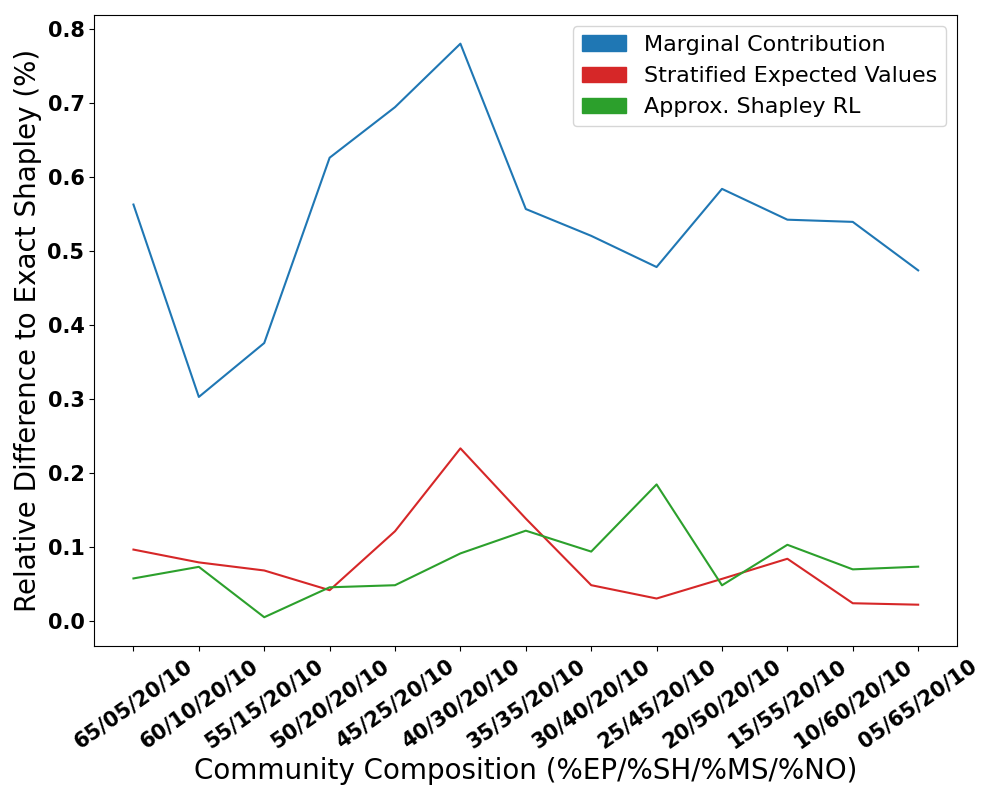}
        \caption{Night Owl}
        \label{fig:4types-ratio-eachTyped}
    \end{subfigure}
    \caption{Individual relative differences to the exact Shapley values for the redistribution methods of four consumer profiles of different community compositions with $N=200$.}
    \label{fig:4types-ratio-eachType}
\end{figure}

Finally, \cref{fig:4types-ratio-eachType} shows the relative differences of each class with changing community composition from \cref{subsec:exp2}. Again, the marginal contribution method shows higher error than the other two redistribution methods. It is also noticeable that the trends of the lines of the marginal contribution and the stratified expected values methods are almost identical between \cref{fig:4types-ratio} and \cref{fig:4types-ratio-eachTyped}, displaying that the error caused by the ``night owl'' class is the main contribution of the overall error of these two methods. Overall, this shows that the precision of the estimation methods is sensitive to more unusual, rarer demand profiles. 

\end{document}